\documentclass[journal]{IEEEtran}

\usepackage{cite}
\usepackage[pdftex]{graphicx}
\usepackage[cmex10]{amsmath}
\usepackage{algorithmic}
\usepackage{array}
\usepackage[caption=false,font=footnotesize]{subfig}
\usepackage{stfloats}
\usepackage{epstopdf}
\usepackage{multirow}
\usepackage{multicol}
\usepackage{amssymb}
\usepackage{amsthm}
\usepackage{setspace}
\usepackage{color}

\begin{document}
\DeclareGraphicsExtensions{.jpg,.pdf,.jpeg,.png,.eps}
\interdisplaylinepenalty=2500
\fnbelowfloat
\newtheorem{definition}{Definition}
\newtheorem{lemma}{\textit{ ~Lemma}}


\title{HetHetNets:\\ Heterogeneous Traffic Distribution in Heterogeneous Wireless Cellular Networks}

\author{Meisam~Mirahsan,~\IEEEmembership{Student Member,~IEEE,}
        Rainer~Schoenen,~\IEEEmembership{Senior Member,~IEEE,}
        and~Halim~Yanikomeroglu,~\IEEEmembership{Senior Member,~IEEE}
\thanks{M. Mirahsan, R. Schoenen, and H. Yanikomeroglu are with the Department of Systems and Computer Engineering, Carleton University, Ottawa, ON, Canada (e-mail: \{mirahsan, rs, halim\}@sce.carleton.ca).}
\thanks{This work is supported in part by Huawei Canada Co., Ltd., and in part by the Ontario Ministry of Economic Development and Innovations ORF-RE (Ontario Research Fund - Research Excellence) program.}
\thanks{Preliminary results from this work were presented, in part, at the IEEE International Conference on Communications (ICC) 2014 Workshop on 5G Technologies, and, in part, at the IEEE Global Communications Conference (Globecom) 2014.}}
\maketitle


\begin{abstract}
A recent approach in modeling and analysis of the supply and demand in heterogeneous wireless cellular networks has been the use of two independent Poisson point processes (PPPs) for the locations of base stations (BSs) and user equipments (UEs). This popular approach has two major shortcomings. First, although the PPP model may be a fitting one for the BS locations, it is less adequate for the UE locations mainly due to the fact that the model is not adjustable (tunable) to represent the severity of the heterogeneity (non-uniformity) in the UE locations. Besides, the independence assumption between the two PPPs does not capture the often-observed correlation between the UE and BS locations.

This paper presents a novel heterogeneous spatial traffic modeling which allows statistical adjustment. Simple and non-parameterized, yet sufficiently accurate, measures for capturing the traffic characteristics in space are introduced. Only two statistical parameters related to the UE distribution, namely, the coefficient of variation (the normalized second-moment), of an appropriately defined inter-UE distance measure, and correlation coefficient (the normalized cross-moment) between UE and BS locations, are adjusted to control the degree of heterogeneity and the bias towards the BS locations, respectively. This model is used in heterogeneous wireless cellular networks (HetNets) to demonstrate the impact of heterogeneous and BS-correlated traffic on the network performance. This network is called HetHetNet since it has two types of heterogeneity: heterogeneity in the infrastructure (supply), and heterogeneity in the spatial traffic distribution (demand).
\end{abstract}


\IEEEpeerreviewmaketitle


\section{Introduction}

\IEEEPARstart{W}{ith} the advent of the increasingly diversified usage scenarios and applications in the envisioned 5G wireless networks, the traffic (demand) in time and space is getting increasingly heterogeneous (non-uniform). Coping with the performance-related challenges in such networks often necessitates the availability of realistic, yet relatively simple and manageable, traffic models. 

The statistics of the signal-to-interference-plus-noise ratio (SINR) are the key to the performance analysis of heterogeneous wireless cellular networks. The signal strengths and interference levels depend strongly on the network geometry, i.e., the relative positions of the transmitters and the receivers. Accordingly, in heterogeneous wireless cellular networks, spatial statistics of the traffic demand (UE distribution) as well as those of the service points (BS distribution) have direct impact on the network performance.

In the traditional single-tier homogeneous wireless cellular networks, usually, the macro-BSs have been assumed to be located in deterministic hexagonal grids \cite{ITU-R2008}. With the advent of multi-tier heterogeneous cellular networks (HetNets)\footnote{In this paper we use the terms HetNet and Heterogeneous Cellular Network (HCN) interchangeably.}, two major transitions have occurred in network modeling. First, small-cell BSs with varying communication characteristics (such as pico-BSs and femto-BSs) have been envisioned to be deployed in the network. It has also become apparent that the regular grid topology assumption for BS locations does not hold anymore mainly due to the fact that small-cell BSs will often be deployed in UE hot-spots which are rather randomly distributed. Therefore, in the contemporary HetNets literature, BS locations have been modeled by Poisson point processes (PPPs).

In addition to the PPP model for the BS locations, it is becoming increasingly common to model the UE locations as another independent homogeneous PPP \cite{andrews2010primer,damnjanovic2011survey}.

Although the recently adopted PPP model for the UE locations is more realistic in comparison to the most common UE location model used in the earlier literature, in which the location of a fixed number of UEs in a cell are determined through two dimensional (2D) uniform random processes, the PPP model does not adequately represent the scenarios in which UEs are heterogeneously distributed (e.g., clustered). Moreover, the independence assumption between the two PPPs (for UE and BS locations) does not capture the correlation between the UE and BS locations observed in reality.

The real UE distributions are seldom pure PPPs. Due to the deployment of small-cell BSs in UE hot-spots, the correlation between UEs and BSs is an apparent phenomenon in HetNets. Network users are usually concentrated at social attractions such as residential and office buildings, shopping malls, and bus stations. Studying the UE distributions of more extreme characteristics and their impact on network performance is thus an important investigation. The requirement is a continuously adjustable and tunable traffic model which can represent the broad possibilities from completely homogeneous cases (e.g., deterministic lattice) to extremely heterogeneous types (e.g., highly clustered scenarios), and from BS-independent UE locations to highly BS-correlated types.

Only when this realistic spatial traffic model is used in HetNet scenarios, the true impact of traffic characteristics on the network performance can be adequately captured. Obviously, the spatial UE distribution has a major impact on the network's key  performance indicators, such as UE rates and outage probabilities which depend directly on UEs' SINR statistics. The spatial UE distribution has also an impact on the network energy efficiency. Especially in HetNets with dense small-cell deployments, the cell switch-off approach \cite{hasan2011green,chen2011fundamental,yanikomeroglu2014novel} is an effective scheme in energy saving; the extent of cell switch-off depends mainly on the distribution of UEs.


\subsection{Contributions of This Paper}

As a first step to fill the explained void in the literature, the contributions of this paper are summarized as follows:
\begin{itemize}
\item A spatial traffic modeling approach with adjustable statistical properties, capturing the severity of the heterogeneity and the extent of the correlation with BSs, is introduced. Although the approach is presented in the context of heterogeneous wireless cellular networks, it is very flexible regarding the underlying network technology and it is general enough to be applied to other contexts as well, including Wi-Fi, ad-hoc, and sensor networks.
\item Mathematical tools in stochastic geometry, including the Voronoi and Delaunay tessellations, are used to illustrate the similarities between traffic modeling in the time and space domains. This similarity leads to the introduction of new geometric inter-point distance measures for capturing the properties of spatial point patterns. These measures are chosen in such a way that they resemble the well-known inter-arrival time (\textit{iat}) in the time domain.
\item Only two parameters are introduced for a fairly accurate description of a heterogeneous and BS-correlated spatial traffic scenario:
\begin{enumerate}
\item The coefficient of variation (CoV) values of the appropriately defined inter-point distance measures are used for specifying the deviations from homogeneity. As stated earlier, the discussed measures in the 2D space domain can be interpreted as the equivalents of the \textit{iat} measure in the one dimensional (1D) time domain. It is worth mentioning that CoV is a commonly used statistic in traffic and queuing theories.
\item The correlation coefficient between the spatial UE distribution and the spatial BS distribution is used for specifying the bias of UEs towards BSs.
\end{enumerate}
\item The developed methodology is demonstrated in a heterogeneous wireless cellular network (HetNet) to illustrate the effects of the realistic traffic modeling on the performance.
\end{itemize}


\subsection{Related Works}

Traffic demand modeling in the time domain has been investigated well in the literature \cite{paxson1995wide, fischer1993markov, heffes1986markov, klemm2003modeling, muscariello2005markov, dainotti2008internet, maheshwari2011joint, xie2012modeling}. Traditionally in voice-only networks, homogeneous Poissonian models were accurate enough to model traffic in time. After the emergence of different applications, such as video and data with variable rates, the Poisson model failed to capture the traffic statistics \cite{paxson1995wide}; as a result, various heterogeneous (super-Poisson) traffic models based on the hidden Markov model (HMM), Markov modulated Poisson process (MMPP) \cite{fischer1993markov} and other stochastic methods have been proposed in the literature and used for performance analysis. 

In the space domain, on the other hand, while there are many papers concentrating on the modeling of base station locations using stochastic geometry \cite{elsawy2013stochastic,andrews2010primer}, there are only few works in the literature which take into account the heterogeneous spatial distribution of traffic demand in wireless cellular networks \cite{bettstetter2007inhomogeneous, qvarfordt2012evaluation, 3GPP, dhillon2012modeling1, taranetz2012performance,dongheon2014,wangload}. To the best of the authors' knowledge, none of the existing works provides a statistically adjustable model representing a variety of possible scenarios for UE distribution.

In \cite{bettstetter2007inhomogeneous}, Bettstetter et al., presented an algorithm to create a random inhomogeneous node distribution based on a neighborhood-dependent thinning approach in a homogeneous PPP. The model, however, can not be used for generating BS-correlated UE patterns, as this is beyond the scope of that model.

In \cite{qvarfordt2012evaluation}, Qvarfordt and Legg presented non-uniform UE layouts (partly clustered around picocells) according to the 3GPP model 4a \cite{3GPP}. However, this model is not designed to adjust the traffic statistically, i.e., the traffic statistics are not measured to be used as an input to the traffic generation function.

Dhillon et al., in \cite{dhillon2012modeling1}, proposed a non-uniform UE distribution model. They start with a higher density of BSs. Then they consider a typical UE located at the origin. After selecting the serving BS, they condition on this active link and independently thin the rest of the BS point process so that the resulting density matches the desired density of the actual BSs. It should be pointed out that the situations in which UEs are clustered, but not necessarily around BSs, are not captured by this method.

The spatial traffic modeling method proposed in \cite{dongheon2014}, by Dongheon Lee et al., suggests that the spatial traffic can be approximated by the log-normal or Weibull distribution. This paper considers the statistics of the spatial traffic distribution, but it does not discuss the modeling of the cross-correlation of traffic with BS locations.

In \cite{Mira1406:Unified} we proposed new measures for capturing traffic characteristics in the space domain. The proposed measures can be considered as the analogues of \textit{iat} in the time domain. Thomas point process was used to generate spatial traffic patterns with desired characteristics. However, the HetNet scenarios are not investigated in \cite{Mira1406:Unified}.

In \cite{mirahsan_gc14} we proposed a novel methodology for the statistical modeling of spatial traffic in wireless cellular networks. The proposed traffic modeling considered the cross-correlation between the UEs and BS locations as well as the CoV as defined in \cite{Mira1406:Unified}. The proposed traffic generation method was a density based method with two phases. First, a BS-biased non-uniform density function for the entire field was generated, then the desired point pattern was produced based on that density function. It should be noted, however, that the generation of the density function for all the points in the field (as required in the method proposed in \cite{mirahsan_gc14}) is computationally intensive. Moreover, the model proposed in \cite{mirahsan_gc14} is not directly applicable on HetNets because the density function is calculated for a homogeneous macro-only scenario.

In this paper, we propose a superior traffic generation method in comparison to that in \cite{mirahsan_gc14} which is also applicable to HetNet scenarios and study the impact of spatially heterogeneous traffic on heterogeneous infrastructure. The proposed method is computationally efficient since it does not require the generation of a density function. Some of the key results in our previous works \cite{Mira1406:Unified} and \cite{mirahsan_gc14} are also presented in this paper in a comprehensive and coherent way with more analytical detail.


\subsection{Organization of the Paper}

The remainder of this paper is organized as follows: In Section \ref{sec:methodology}, our traffic modeling methodology is presented. In Section \ref{sec:measurement}, new traffic measures and appropriate statistics are introduced. The traffic generation process is described in Section \ref{sec:generation}. The numerical results for traffic modeling and network performance analysis are presented in Section \ref{sec:results}, and the concluding remarks are made in Section \ref{sec:conclusion}.


\section{Traffic Modeling Methodology}
\label{sec:methodology}

Heterogeneous wireless cellular network models receive traffic patterns in the time domain as their input for performance analysis. The key performance indicators such as, average user data rates, system spectral efficiency, and outage probability, are calculated as the system outputs. The traffic patterns are usually generated by random processes (e.g., PPP, MMPP, HMM, ...), which are called traffic generators (TGs).

TGs, in turn, receive a number of input parameters (TGIPs) to generate traffic patterns with various statistical properties. Therefore, a main track of research in traffic modeling has been to fit TGIPs to generate traffic patterns with the desired statistical properties (e.g., measured statistics from a given real traffic trace) \cite{andersen1998markovian, casale2008interarrival, xie2012modeling}.

\begin{figure*}
\centering
\includegraphics[width=\textwidth]{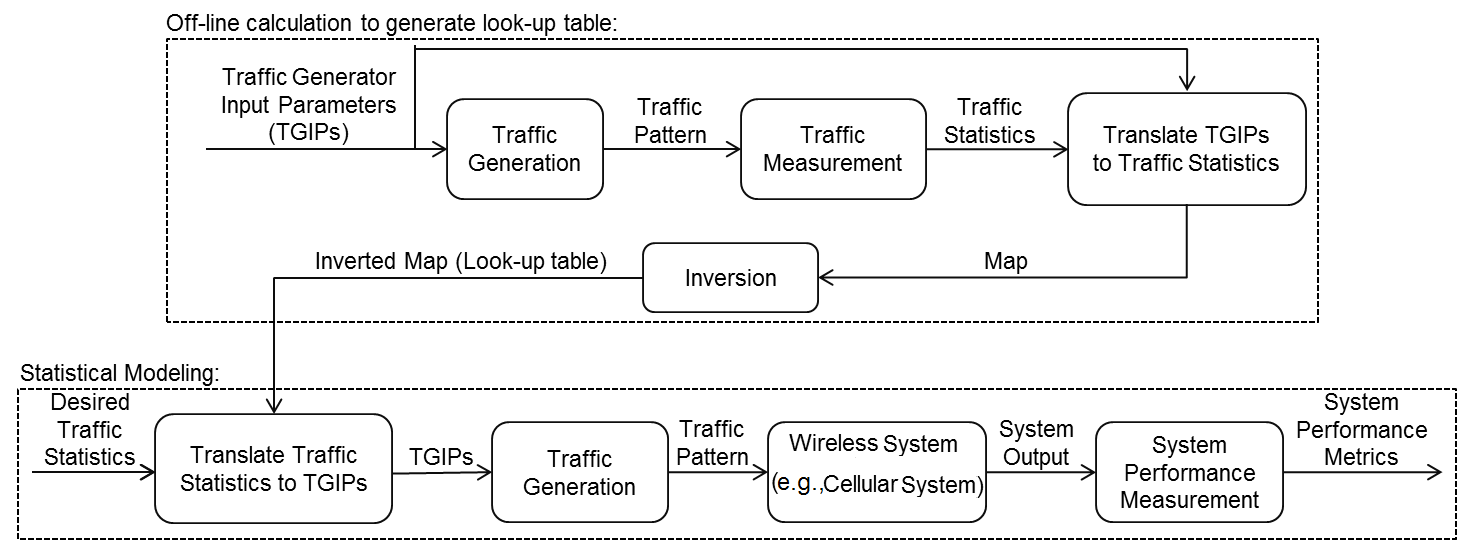}
\caption{Modeling procedure: The desired statistical properties of traffic as the modeling inputs are translated to the appropriate TGIPs, and traffic with known characteristics is generated to be used for network performance analysis (lower dashed box). The look-up table for translation is generated in advance via off-line calculations (upper dashed box) \cite{mirahsan_gc14}.}
\label{fig:modelingprocedure}
\end{figure*}

Figure \ref{fig:modelingprocedure} demonstrates our proposed traffic modeling methodology in the space domain. A somewhat similar methodology has been used in the literature for traffic modeling in the time domain \cite{xie2012modeling}.

The proposed methodology is described as follows:
\begin{itemize}
\item The desired statistical properties of traffic as the modeling inputs are fed to the traffic generator.
\item Since the traffic generator expects TGIPs, the desired statistics are first translated to the appropriate TGIPs. \item Traffic with known characteristics is then generated to be used for network performance analysis.
\item The translation look-up table is developed off-line in advance:
\begin{itemize}
\item The entire range of the feasible TGIP values are fed to the traffic generator and the traffic pattern statistics are measured at the output. Therefore, a complete map from TGIP values to the corresponding traffic statistics is produced.
\item The conversion function from TGIPs to the traffic statistics is obtained by fitting methods.
\item The look-up table from traffic statistics to the corresponding TGIPs can be prepared by inversion of the map or the function from TGIPs to the measured statistical properties.
\end{itemize}
\end{itemize}
A mathematical description of the process is given in Section \ref{subsec:improvedmethod}.

The method described above is capable of generating a wide range of heterogeneity possibilities. However, an appropriate next-step is to tune the model according to the real world measurements and to extract the accurate model parameters for generating realistic traffic patterns. In \cite{meisammap2015}, we used the maps of Paris, France, obtained from OpenStreetMaps \cite{OSMorg}, to study the spatial traffic heterogeneity of outdoor users in the denser areas of the city center.


\section{Traffic Measurement}
\label{sec:measurement}

A perfect model will require all the statistics of the traffic measures for the generated traffic patterns and those for the real traffic traces, including the cumulative distribution functions (CDFs) and auto-correlation functions, to match. Therefore, a perfect match is not practical as this requires the use of extremely complicated models with a very large number of parameters. As a result, simplified models are commonly used which only consider the first few moments of the traffic. In the time domain, usually the mean, CoV, and auto-correlation, and rarely the third moment, are considered to match with those of the real traffic trace. 

For the measurement of traffic patterns, first, an appropriate measure must be selected to capture the traffic properties. In the time domain, \textit{iat} is the most popular and well accepted measure.

In Section \ref{subsec:measures}, the equivalent measures for the statistics of the spatial traffic are introduced. The spatial traffic statistics are described in Section \ref{subsec:statistics}.


\subsection{Traffic Measures}
\label{subsec:measures}

Packet arrivals in the time domain can be modeled by a one-dimensional (1D) point process. A fixed \textit{iat} between packets generates the maximum homogeneity (deterministic lattice). An exponentially distributed \textit{iat} generates complete randomness (PPP in 1D). For generating sub-Poisson patterns (patterns with more homogeneity than Poisson), one way is to generate a perfect lattice, then apply a random displacement (perturbation) on its points \cite{rataj1993convergence,lucarini2008symmetry}. Various models for generating super-Poisson patterns (patterns with more heterogeneity than Poisson) have been proposed in the literature which are mostly based on hierarchical randomness and Markov models \cite{paxson1995wide, dainotti2008internet, xie2012modeling}.

A 1D point pattern in the time domain can be measured mathematically in many different ways. One may use the interval count, $N(a, b] = N_{b} - N_{a}$ ($N_{t}$ representing the number of points arrived before time $t$), which is a density-based measure and divides the whole domain into smaller windows and counts the number of pattern points in each window. A disadvantage of the density-based measures is that they are parameterized by the window size. Finding an appropriate window size is itself a challenging question and does not have a unique answer for all applications.

The inter-arrival time, \textit{iat}, $I_{i} = T_{i+1} - T_{i}$ ($T_{i}$ representing the time of arrival of point $i$), is the most popular and best accepted measure because it is distance-based rather than density-based, and it considers the distance between every two neighboring points in the domain. The CoV for \textit{iat}s is defined as
\begin{equation}
C_{_{I}} = \frac{\sigma_{_{I}}}{\mu_{_{I}}},
\end{equation}
where $\mu_{_{I}} > 0$ and $\sigma_{_{I}}$ are the mean and standard deviation of \textit{iat}s, respectively (note that $\mu_I > 0$ if $t > 0$). For a perfect 1D lattice, the constant \textit{iat} has $C_{_{I}} = 0$. For a 1D Poisson pattern, $C_{_{I}}=1$, since for an exponential distribution with parameter $\lambda$, the standard deviation and the mean are both $\mu_{_{I}} = \sigma_{_{I}} = \lambda$. Sub-Poisson processes have $0 < C_{_{I}} < 1$ and super-Poisson processes have $C_{_{I}} > 1$.

The UE locations in a heterogeneous wireless cellular network in space domain can be modeled by a two-dimensional (2D) or three-dimensional (3D) point process. A very inclusive review of Point processes in space domain is conducted in \cite{blaszczyszyn2012clustering}. A fixed distance between points generates perfect homogeneity (deterministic lattice). On the contrary, the Poisson distribution generates complete randomness. For generating sub-Poisson patterns, one way is to generate a perfect lattice, then apply a random perturbation on its points \cite{rataj1993convergence,lucarini2008symmetry,blaszczyszyn2012clustering}.

As mentioned above, in time domain, the distance-based measure \textit{iat} captures heterogeneity by one non-parameterized real value $C_{_{I}}$. In multi dimensions, however, there is no natural ordering of the points, so finding the analogue of the \textit{iat} is not straightforward. There are many density-based heterogeneity measures in the literature such as Ripley's K-function and pair correlation function \cite{blaszczyszyn2012clustering}, but they are all parameterized. For introducing distance-based measures, there is the problem of defining the 'next point' or the 'neighboring points' in multi-dimensional domains.

The first and simplest candidate for characterizing a neighboring point in a multi-dimensional domain is to consider the nearest-neighbor. This leads to the nearest-neighbor distance measure \cite{clark1954distance}. However, the nearest-neighbor distance measure in 1D time domain is not the analogue of the \textit{iat} because it considers the $\min \{I_{i},I_{i+1}\}$ for every point $T_{i}$. It is shown in Section \ref{sec:results} in our simulation results that the nearest-neighbor distance fails to capture the process statistics in multi-dimensional domains because this measure ignores the neighbors other than the closest one. The next candidate is the distance to the $k^{th}$ neighbor. However, determining $k$ globally is not possible because every point may have a different number of neighbors.

In the following, novel distance-based and non-parameterized measures in the space domain are proposed based on the Voronoi and Delaunay tessellations.

\begin{definition}
Voronoi Tessellation \cite[p.~1]{barr2008applications}: Given a point pattern $P = \{p_{1}, p_{2}, ... , p_{n}\}$ in d-dimensional space $\mathbb{R}^{d}$, the Voronoi tessellation $VT = \{c_{p_{1}},c_{p_{2}}, ... , c_{p_{n}}\}$ is the set of cells such that every location, $y \in c_{p_{i}}$, is closer to $p_{i}$ than any other point in $P$ . This can be expressed formally as
\begin{equation}
c_{p_{i}} = \left \{  y \in \mathbb{R}^{d} : \left | y - p_{i} \right | \leq \left | y - p_{j} \right | \ \textrm{for} \ i,j \in 1,...,n \right \}.
\end{equation}
\end{definition}

\begin{definition}
Delaunay Tessellation \cite[p.~11]{moller2007stochastic}: The Voronoi tessellation in $\mathbb{R}^{d}$ has the property that each of its vertices is given by the intersection of exactly $d + 1$ Voronoi cells. The corresponding $d + 1$ points define a Delaunay cell. So the two tessellations are said to be dual.
\end{definition}

\begin{figure}
   \centering
   \includegraphics[width=0.5\columnwidth]{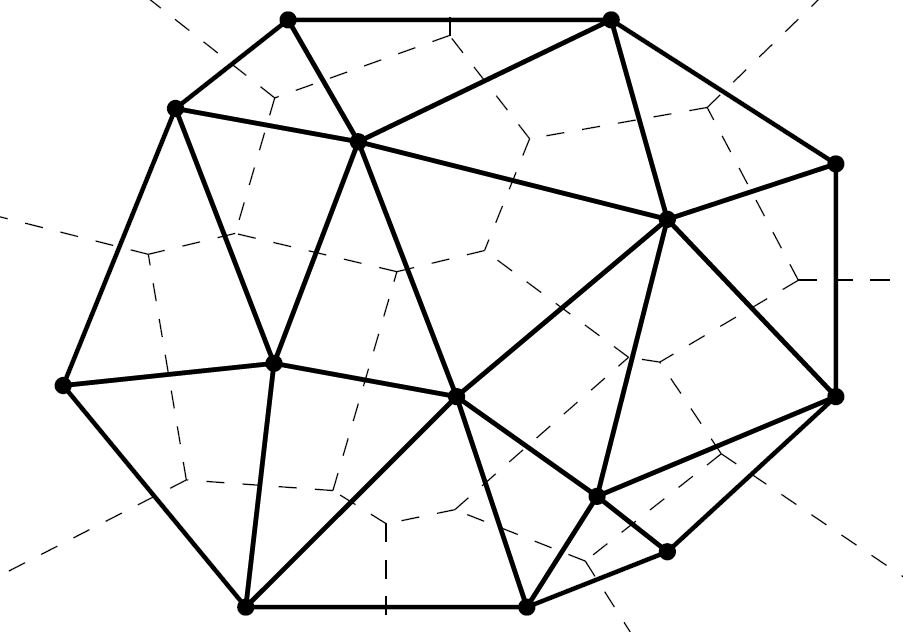}
   \caption{Voronoi and Delaunay tessellations: For a point pattern represented by bullets, the associated Voronoi tessellations (dashed lines) and Delaunay tessellations (solid lines) are illustrated.}
   \label{fig:voronoidelaunay}
\end{figure}

Figure \ref{fig:voronoidelaunay} demonstrates a pattern of points with its Voronoi tessellation (dashed lines) and Delaunay tessellations (solid lines).

\begin{definition}
Natural Neighbor \cite[p.~3]{sirovich2002natural}: Every two points sharing a common edge in Voronoi tessellation or equivalently every two connected points in Delaunay tessellation of a point process are called 'natural neighbors'.
\end{definition}

Natural neighborship gives an inspiration of neighboring relation in multi-dimensional domains and leads us to the analogues of the well accepted \textit{iat} measure in multi dimensions. Various statistical inferences based on different properties of cells generated by these tessellations can be considered for the measurement of a point pattern.

The Voronoi cell area or Voronoi cell volume $V$ is the first natural choice. For a lattice process, all the cell areas in 2D or cell volumes in 3D are equal which results in $C_{_{V}}= 0$. The statistics of the Voronoi cells for a PPP (Poisson-Voronoi tessellation) are well investigated in the literature \cite{barr2008applications,moller2007modern,gilbert1962random,borovkov2007simulation,tanemura2003statistical}. Square-rooted Voronoi cell area in 2D or cube-rooted Voronoi cell volume in 3D can also be considered. We are interested in Voronoi cell area $V$ because it can be considered as an analogue of \textit{iat} in the time domain.

The main focus of this paper is the spatial aspect of traffic distribution because the temporal aspect is well-investigated in the literature. The combination of the temporal domain and the spatial domain, which is a very challenging problem, can be considered as an extension of this paper. However, it is critical to understand the heterogeneity in the space domain thoroughly before proceeding to the combined domain. In that case, some other metrics such as the inverted Voronoi cell area, $1/V$, could be used to normalize the user target rates in the time domain. Therefore, a combined metric such as $R/V$ can be of interest in the combined domains; this remains as an interesting future work.

The next proposed measure is the Delaunay edge length $E$. The statistics of Delaunay tessellations is investigated in \cite{muche1996distributional,rathie1992volume,miles1972random}. The mean value of the lengths of Delaunay edges of every point can also be considered.

A Delaunay tessellation divides the space into triangles or tetrahedrons in 2D and 3D, respectively. The area distribution of the triangles or the volume distribution of tetrahedrons can determine the properties of the underlying pattern. 

The Voronoi and Delaunay tessellations can be applied on a 1D process which models traffic in time domain. In this case, the introduced distance-based measures are converted to time domain measures. The basic statistics of these measures for a PPP in one, two, and three dimensions, and their analogues in time domain, are summarized in Table \ref{tab:distmeasures}.

\begin{table*}
\centering
\caption{\normalfont{Basic statistics of distance-based measures for a PPP in one, two and three dimensions and their analogues in time domain: $i$ is the process point index, $\lambda$ is the exponential distribution parameter for inter-arrival time and $\Lambda$ is the mean intensity of point processes\cite{Mira1406:Unified}.}}
\label{tab:distmeasures}
    \begin{tabular}{|l|l|l|l|l|l|}
    \hline
    Distance-based measures & Time domain & Statistics & 1D & 2D & 3D \\ \hline

    \multirow{3}{*}{Nearest-neighbor distance (G)} &  \multirow{3}{*}{$\min \{I_{i},I_{i+1}\}$}
		& Mean ($\mu$) & $0.5 \lambda^{-1}$ & $0.5 \Lambda^{-0.5}$ & $0.5539 \Lambda^{-0.33}$\\ 
		& & Variance ($\sigma^{2}$) & $0.25 \lambda^{-2}$ & $0.0683 \Lambda^{-1}$ & $0.04 \Lambda^{-0.66}$\\
      & & CoV ($C$) & 1 & 0.653 & 0.364\\ \hline

    \multirow{3}{*}{Voronoi cell area/volume (V)}    &  \multirow{3}{*}{$\frac{I_{i}+I_{i+1}}{2}$}   
       & Mean ($\mu$) & $\lambda^{-1}$ & $\Lambda^{-1}$ & $\Lambda^{-1}$\\
       & & Variance ($\sigma^{2}$) & $\lambda^{-2}$ & $0.28 \Lambda^{-2}$ & $0.18 \Lambda^{-2}$\\
       & & CoV ($C$) & 1 & 0.529 & 0.424\\ \hline

    \multirow{3}{*}{Delaunay cell edge length (E)}  &  \multirow{3}{*}{$I_{i}$} 
		& Mean ($\mu$) & $\lambda^{-1}$ & $1.131 \Lambda^{-0.5}$ & $1.237 \Lambda^{-0.33}$\\
      & & Variance ($\sigma^{2}$) & $\lambda^{-2}$ & $0.31 \Lambda^{-1}$ & $0.185 \Lambda^{-0.66}$\\
      & & CoV ($C$) & 1 & 0.492 & 0.347\\ \hline

    \end{tabular}
\end{table*}

\begin{figure}
   \centering
   \subfloat{\includegraphics[width=\columnwidth]{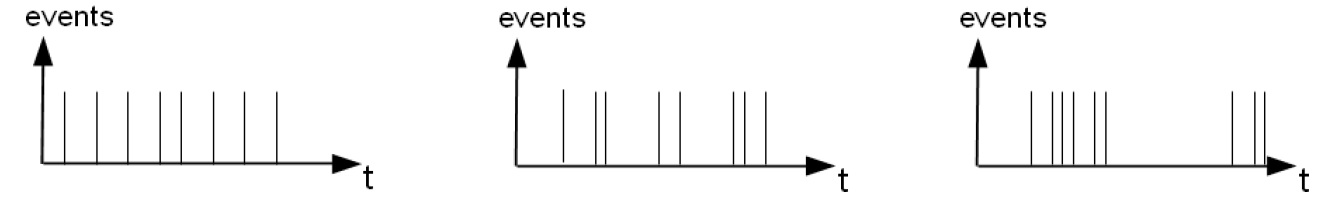}}
   \vfill
   \subfloat{\includegraphics[width=0.32\columnwidth]{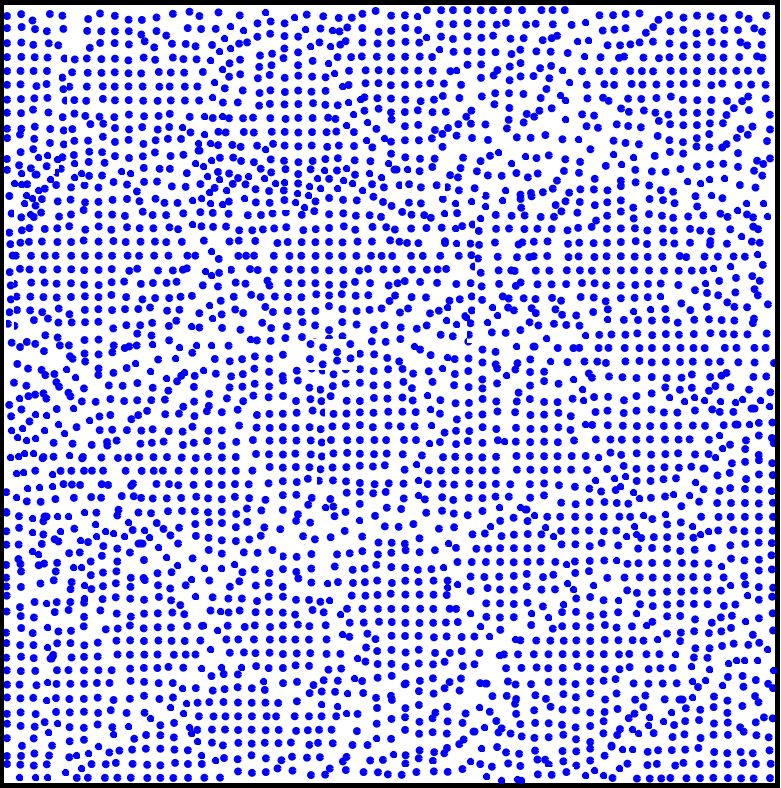}}
   \hfill
   \subfloat{\includegraphics[width=0.32\columnwidth]{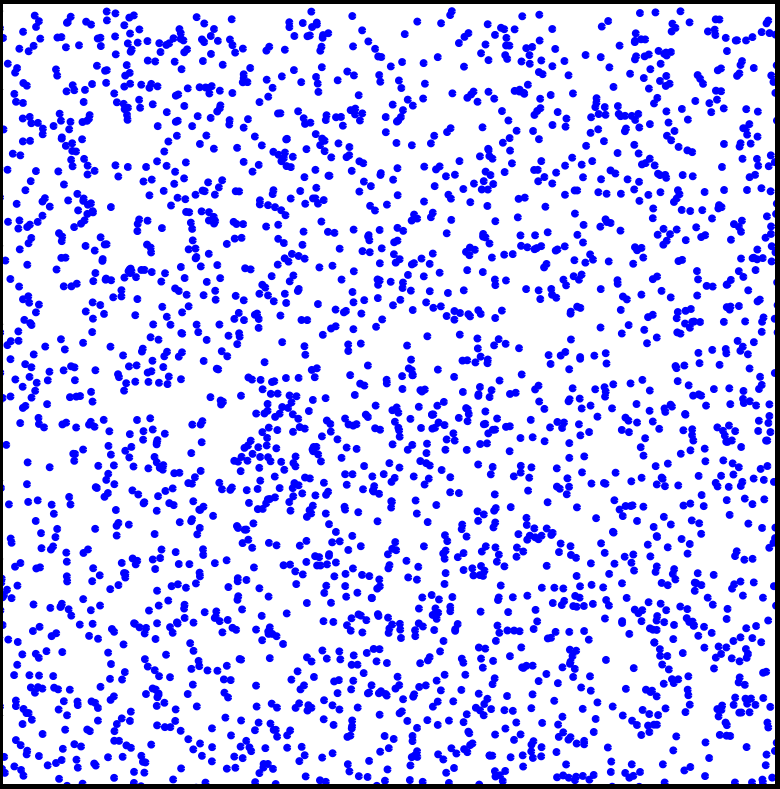}}
   \hfill
   \subfloat{\includegraphics[width=0.33\columnwidth]{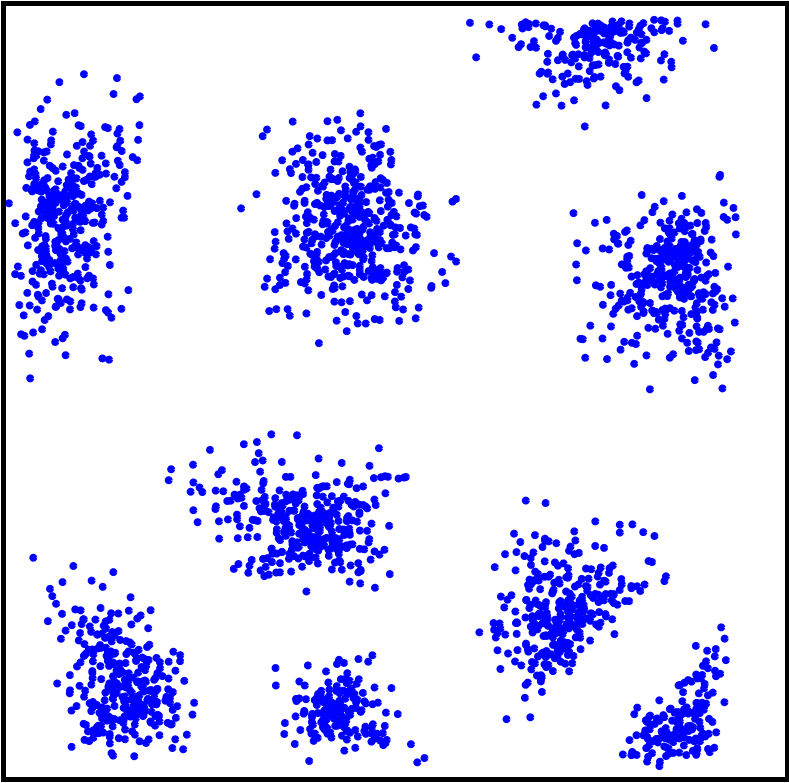}}
   \hfill
   \caption{Domain similarity: Realizations of processes with sub-Poisson ($0 < C < 1$), Poisson ($C=1$), and super-Poisson ($ C > 1$) characteristics, respectively, from left to right in time domain (top) and space domain (bottom).}
   \label{fig:domainsimilarity}
\end{figure}

In order to use the above mentioned measures as an analogue of \textit{iat}, one needs to normalize their CoV with the CoV values of \textit{iat} in the time domain. For the complete homogeneity case, the CoV values are already zero, same as \textit{iat} in the time domain. To normalize the CoV values of the complete random case to $1$, it is required to divide the measure by the values presented in Table \ref{tab:distmeasures}. Figure \ref{fig:domainsimilarity} demonstrates realizations of processes with sub-Poisson, Poisson, and super-Poisson characteristics.

\subsection{Traffic Statistics}
\label{subsec:statistics}

Assuming that UEs have the same altitude, spatial traffic is modeled as a 2D point pattern $U \in \mathbb{R}^{2}$ which is generated by a generator point process $\Phi_{_{U}}$. In this paper, the mean, $\mu_{_{m}}$, and the CoV, $C_{_{m}} = \sigma_{_{m}}/\mu_{_{m}}$, are the desired statistics of traffic where $m$ is the traffic measure and $\sigma_{_{m}}$ is the standard deviation of $m$ (the third-moment and auto-correlation are stated as future extensions in Section \ref{sec:conclusion}).

\begin{figure}
\centering
\subfloat[BS weighted-Voronoi tessellation (see \textbf{Acknowledgment})\label{fig:bs_voronoi}]{
      \includegraphics[width=0.45\columnwidth]{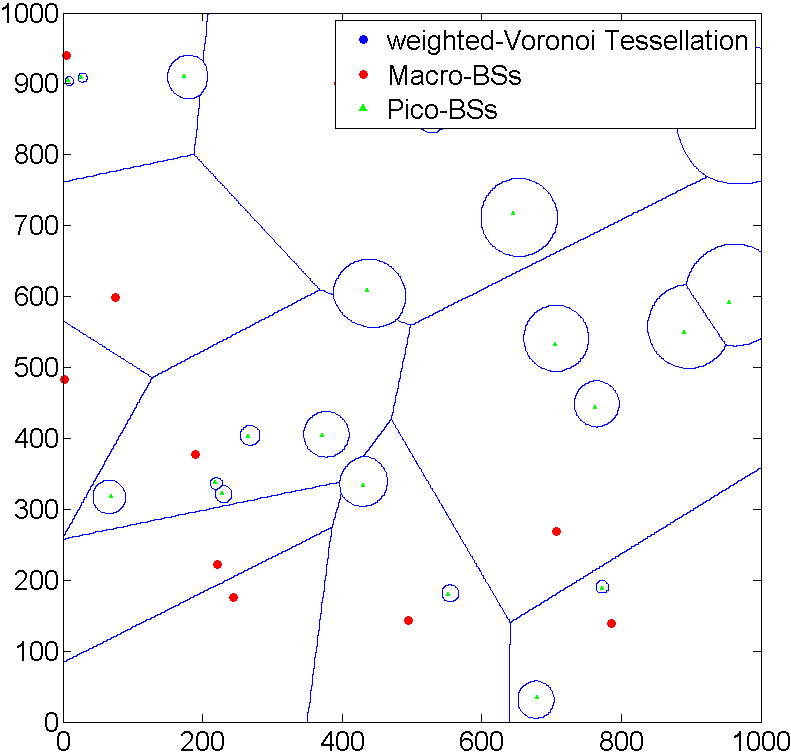}}
\subfloat[potential distribution\label{fig:potentials}]{
      \includegraphics[width=0.45\columnwidth]{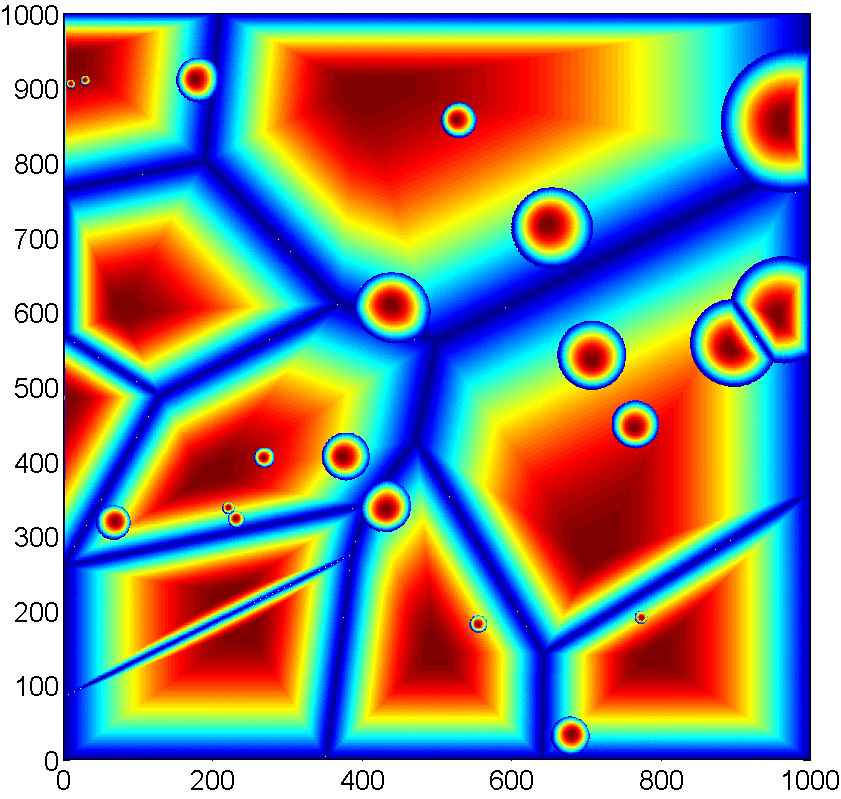}}
\caption{The weighted-Voronoi tessellation of BSs divides the whole field into cells associated with each BS consisting of all points closer, in terms of received signal power, to that BS than to any other BS. Every point in the field is then associated with a potential value between $-1$ and $+1$, according to (\ref{eq:potential}).}
\label{fig:infra}
\end{figure}

Along with the mean and the CoV, which capture the heterogeneity of traffic, a very important statistic of traffic in space which affects the network performance is the bias of the UE distribution to the BS distribution (i.e., the correlation between the two distributions). BS locations can be modeled by a superposed 2D point pattern $B = \bigcup_{k=1}^{K} B_k, B_k \in \mathbb{R}^{2}$, where $K$ is the number of tiers of BSs in heterogeneous infrastructures and $B_k$ is the set of BSs of type $k$ operating with power $P_k$ and generated by a point process $\Phi_{_{k}}$ with density $\lambda_k$. The weighted-Voronoi tessellation of BSs divides the entire field into Voronoi cells associated with each BS consisting of the area closer, in terms of received signal power, to that BS than to any other BS. A sample realization is shown in Fig. \ref{fig:infra}(a).

To measure the joint distribution of UEs and BSs, we define the following potential function:
\begin{definition}
Potential Function: Every point $(x,y)$ in the field is associated with a potential value $P(x,y) \in \left [ -1, +1 \right ]$. The $P$ function must have the following properties:
\begin{enumerate}
\item $P(x,y) = +1$, for cell center points, 
\item $P(x,y) = -1$, for Voronoi cell edge points,
\item $\iint_{A_{i}}P(x,y)dxdy = 0, \;\; \forall i$,
\end{enumerate}
where $i$ is the index for BSs and $A_{i}$ is the Voronoi cell area associated with BS $B_{i}$.
\end{definition} 

The first property assures that if a UE is at the cell center, its correlation with BSs is measured as $+1$. The second property assures that if a UE is at the cell edge, its correlation with BSs is measured as $-1$. The third property assures that if all UEs are distributed homogeneously (independent from BSs) in the area, their mean cross-correlation with BSs is equal to $0$.

One may identify many functions which satisfy the requirements of the potential function defined above. In this paper, we consider the simplest polynomial function.

\begin{lemma}
The simplest polynomial function which satisfies the requirements of the potential function is
\begin{equation}
\label{eq:potential}
P(x,y) = \frac{-2(d(x,y))^{2}}{(D(x,y))^{2}}+1,
\end{equation}
where $d(x,y)$ is the distance of the point $(x,y)$ to the associated cell center, and $D(x,y)$ is the length of the line connecting the associated cell center to the associated Voronoi cell edge through point $(x,y)$ as shown in Fig. \ref{fig:Dd}.
\end{lemma}
\textbf{\textit{~~Proof.}} See \textbf{Appendix.}

Figure \ref{fig:infra}(b) illustrates the potential distribution associated with the Voronoi tessellation in Fig. \ref{fig:infra}(a).

\begin{figure}
\centering
\includegraphics[width=0.5\columnwidth]{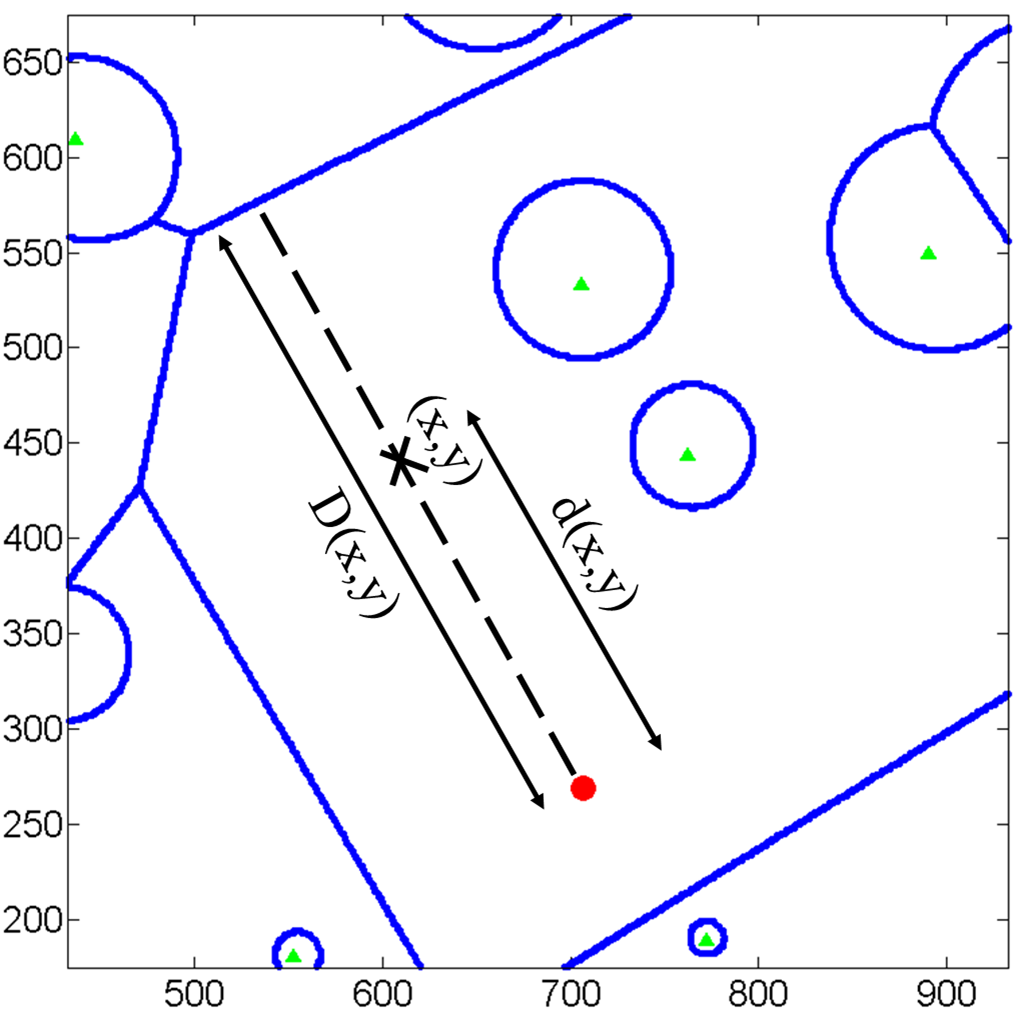}
\caption{The $d(x,y)$ is the distance of the point $(x,y)$ to the associated cell center, and $D(x,y)$ is the length of the line connecting the associated cell center to the associated Voronoi cell edge through point $(x,y)$.}
\label{fig:Dd}
\end{figure}

Using this potential function, the joint moments
\begin{equation}
\mathbb{E}\left [ P^{i}\Lambda^{j} \right ], \; i,j \ge 0,
\end{equation}
and joint central moments
\begin{equation}
\mathbb{E}\left [ (P-\mu_{_{P}})^{i} (\Lambda-\mu_{_{\Lambda}})^{j} \right ], i,j \ge 0,
\end{equation}
can be calculated, where $\mu_{_{P}}$ is the mean value of $P$, $\mu_{_{\Lambda}}$ is the mean value of UE density $\Lambda$, and $\mathbb{E} \left [ x \right ]$ is the expected value of $x$ \cite{alberto1994probability}. To have a normalized measure, we use correlation coefficient which is defined as
\begin{equation}
\rho = \frac{\sigma_{_{P\Lambda}}}{\sigma_{_{P}}\sigma_{_{\Lambda}}},
\end{equation}
where
\begin{equation}
\sigma_{_{P\Lambda}} = \mathbb{E}\left [ (P-\mu_{_{P}})(\Lambda-\mu_{_{\Lambda}}) \right ]
\end{equation}
is the covariance of $P$ and $\Lambda$, $\sigma_{_{P}}$ is the standard deviation of $P$, and $\sigma_{_{\Lambda}}$ is the standard deviation of $\Lambda$. So, $\rho$ can be defined as
\begin{equation}
\rho = \frac{\iint ( \Lambda(x,y)-\mu_{\Lambda})(P(x,y)-\mu_{P})dxdy}{\sqrt{(\iint( \Lambda(x,y)-\mu_{\Lambda})^{2}dxdy)(\iint(P(x,y)-\mu_{P})^{2}dxdy)}}.
\end{equation}

For a UE pattern $U$ (a realization of $\Phi_{_{U}}$), the correlation coefficient is calculated as
\begin{equation}
\label{eq:rho}
\rho = \frac{\sum_{u \in U} P_{u}}{\left | U \right |},
\end{equation}
where $P_{u}$ is the potential value at point $u$ and $\left | U \right |$ is the number of points in $U$. A pattern with $\rho = +1$ means that all UEs have gathered at the cell centers, a pattern with $\rho = 0$ means that the UE distribution is independent from the BS distribution, and a pattern with $\rho = -1$ means that all UEs have gathered at the cell edges.


\section{Heterogeneous Traffic Generation}
\label{sec:generation}

A basic traffic generation method is presented in Section \ref{subsec:basicmethod}, which is further improved in Section \ref{subsec:improvedmethod}.


\subsection{The Basic Method}
\label{subsec:basicmethod}

UEs in a heterogeneous wireless cellular network are attracted to social attractors (SA) such as buildings, bus stations, shopping centers, and other social places. Furthermore, with the proliferation of small-cells and the deployment of BSs near potential SAs, the distance between SAs and BSs is decreasing. The requirement is a traffic generation method which is able to generate most of the possible traffic distributions including the following:
\begin{itemize}
\item situations in which UEs are not attracted to SAs at all and are completely random, while SAs in turn are distributed independently from BSs,
\item situations in which UEs are not attracted to SAs at all and are completely random, while SAs are close to BSs,
\item situations in which UEs are highly attracted to SAs, while SAs are distributed independently from BSs,
\item situations in which UEs are highly attracted to SAs, and SAs are close to BSs.
\end{itemize}
Covering the entire range of the above discussed cases, our proposed traffic generation method is described as follows. Obviously, the proposed method is just one way of generating adjustable heterogeneous and BS-correlated traffic; i.e., it is not the only way. More accurate methods can be developed based on the real world data analysis.

Starting with a Poissonian distribution of BSs ($B$), including macro-BSs, pico-BSs, and femto-BSs, an independent Poissonian distribution of SAs ($S$), and an independent distribution of UEs ($U$), we move every SA ($\vec{S}_{i}$) towards its closest BS, in terms of the received signal power ($\vec{B}_{S_{i}}$), by a factor of $\alpha \in \left[0, 1\right]$; so the SA's new location ($\vec{S}^{new}_{i}$) is calculated as
\begin{equation}
\vec{S}^{new}_{i} = \alpha \vec{B}_{S_{i}} + (1-\alpha) \vec{S}_{i}.
\end{equation}

Then we move every UE ($\vec{U}_{i}$) towards its closest SA, in terms of the Euclidean distance ($\vec{S}_{U_{i}}$), by a factor of $\beta \in \left[0 \; 1\right ]$; so the UE's new location ($\vec{U}^{new}_{i}$) is calculated as
\begin{equation}
\vec{U}^{new}_{i} = \beta \vec{S}_{U_{i}} + (1-\beta) \vec{U}_{i}.
\end{equation}

The initial distribution of UEs can be Poissonian ($C=1$) or deterministic ($C=0$). In this paper, to generate sub-Poisson traffic ($C<1$), we start with a deterministic lattice.

The proposed method is general enough to generate traffic with negative bias to BSs, i.e., UEs gathering at cell edges. To generate this type of traffic, UEs must be moved to cell edges instead of cell centers.


\subsection{The Enhanced Method}
\label{subsec:improvedmethod}

\begin{figure}
   \centering
   \subfloat[$\beta$ is fixed and deterministic]{\includegraphics[width=0.5\columnwidth]{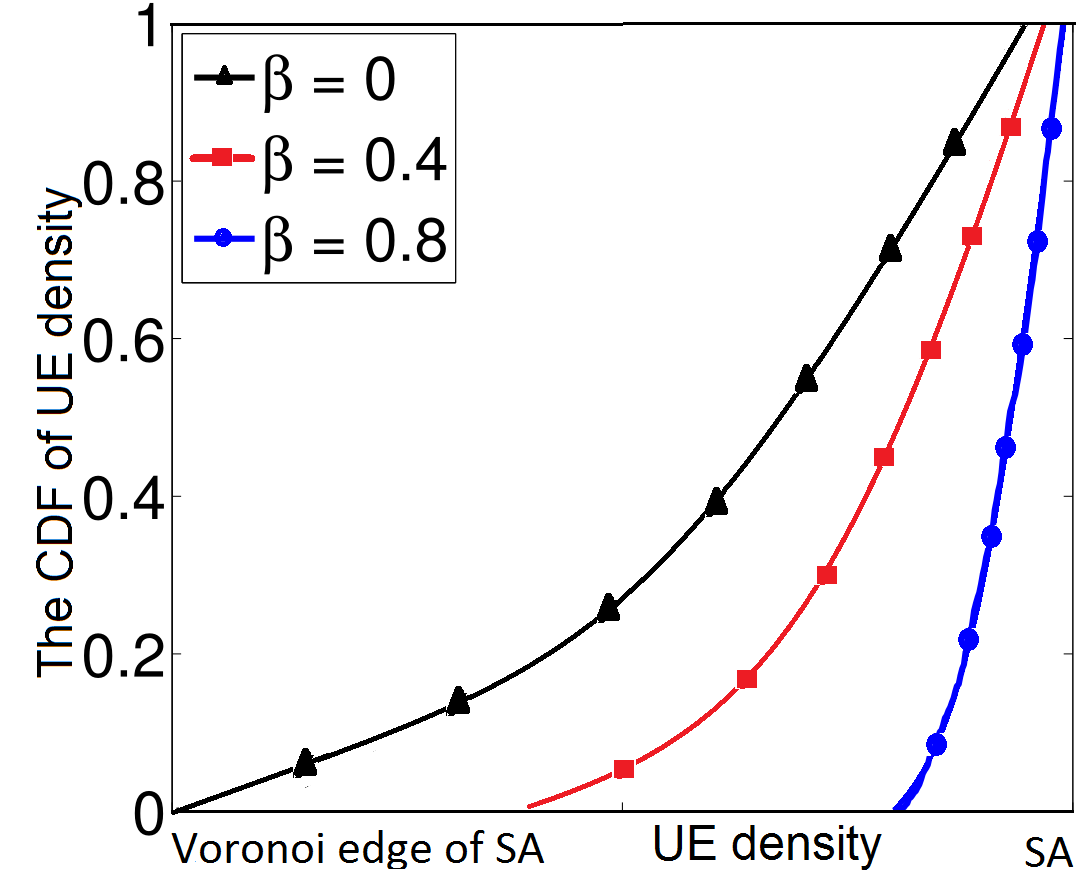}\label{fig:fixcdf}}
   \hfill
   \subfloat[$\beta$ is random (normal distribution)]{\includegraphics[width=0.5\columnwidth]{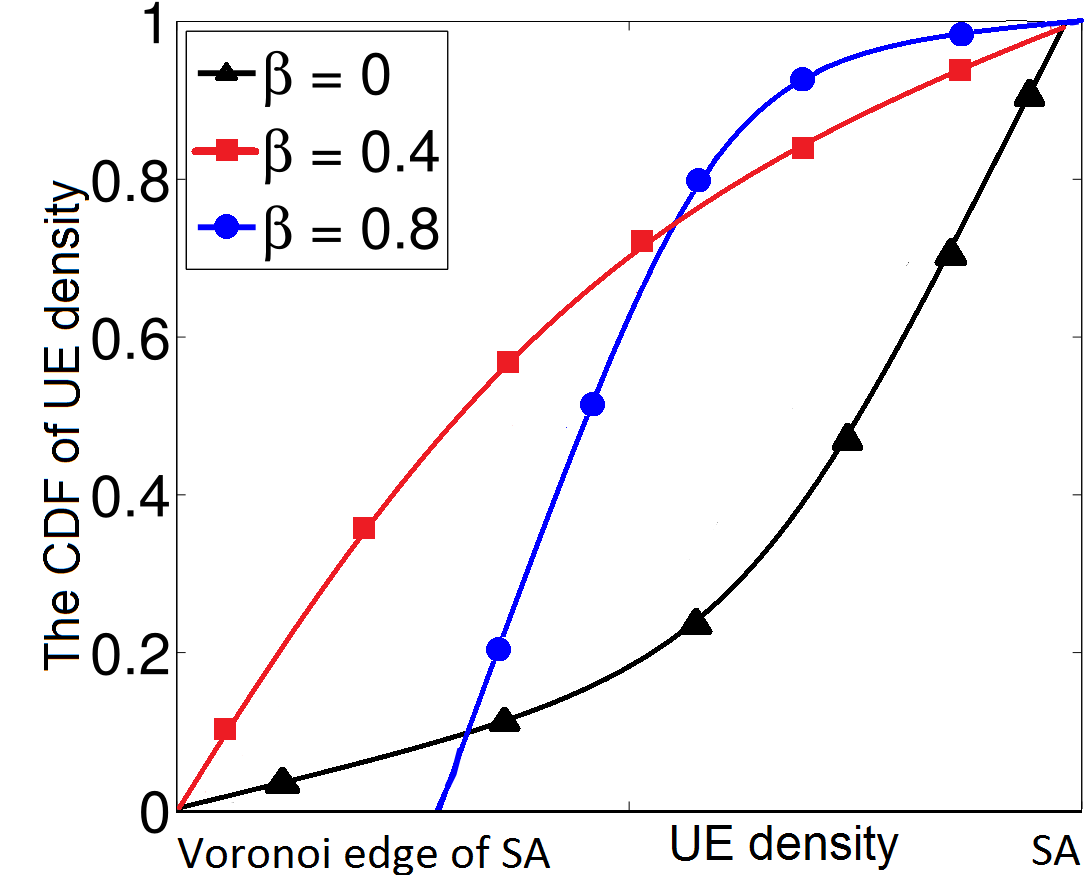}\label{fig:rndcdf}}
   \caption{The CDF of the distribution of UEs in the Voronoi cell area of SAs: (a) When $\beta$ is deterministic, the UEs are moved towards SAs but the UE distribution shape is fixed and the Voronoi cell edge area of SAs remains empty with no UEs. (b) When $\beta$ is a random variable with normal distribution, even for high $\beta$ values, the Voronoi cell edges are not empty and there is a low probability for UEs existing at cell edges.}
   \label{fig:cdfs}
\end{figure}

An undesired property of the basic method described in \ref{subsec:basicmethod} is that when UEs move towards the SAs, some areas of the network become empty having no UEs. In other words, the UE distribution shape stays unchanged and it only shrinks to a smaller area. Figure \ref{fig:cdfs}(a) shows the CDF of the UE density of network after moving UEs towards SAs with different $\beta$ values using the basic method. Even for small values of $\beta$ (e.g., $\beta$=0.4), the SA Voronoi cell edge area has no UEs.

To resolve this undesired feature of the basic method, we introduce an enhanced method as follows. Instead of having a fixed value for $\beta$, in the enhanced method we model $\beta$ as a random variable with
\begin{equation}
\label{eq:beta}
\beta \sim \text{N}(\mu_{\beta},\sigma_{\beta}),
\end{equation}
where the mean $\mu_{\beta} \in \left [0, \; 1 \right]$ indicates the closeness factor of UEs to SAs (an accurate model based on the real world measurements may adopt another distribution for $\beta$, however, since we don’t have access to real traffic measurements, in this paper we choose normal distribution). In (\ref{eq:beta}), the $\sigma_{_{\beta}}$ value should have the following characteristics:
\begin{itemize}
\item $\beta \rightarrow 0 \Rightarrow \sigma_{_{\beta}} \rightarrow 0$,
\item $\beta \rightarrow 1 \Rightarrow \sigma_{_{\beta}} \rightarrow 0$,
\item $\beta \rightarrow 0.5 \Rightarrow \sigma_{_{\beta}}$ is maximized.
\end{itemize}

\begin{figure}
\centering
\includegraphics[width=\columnwidth]{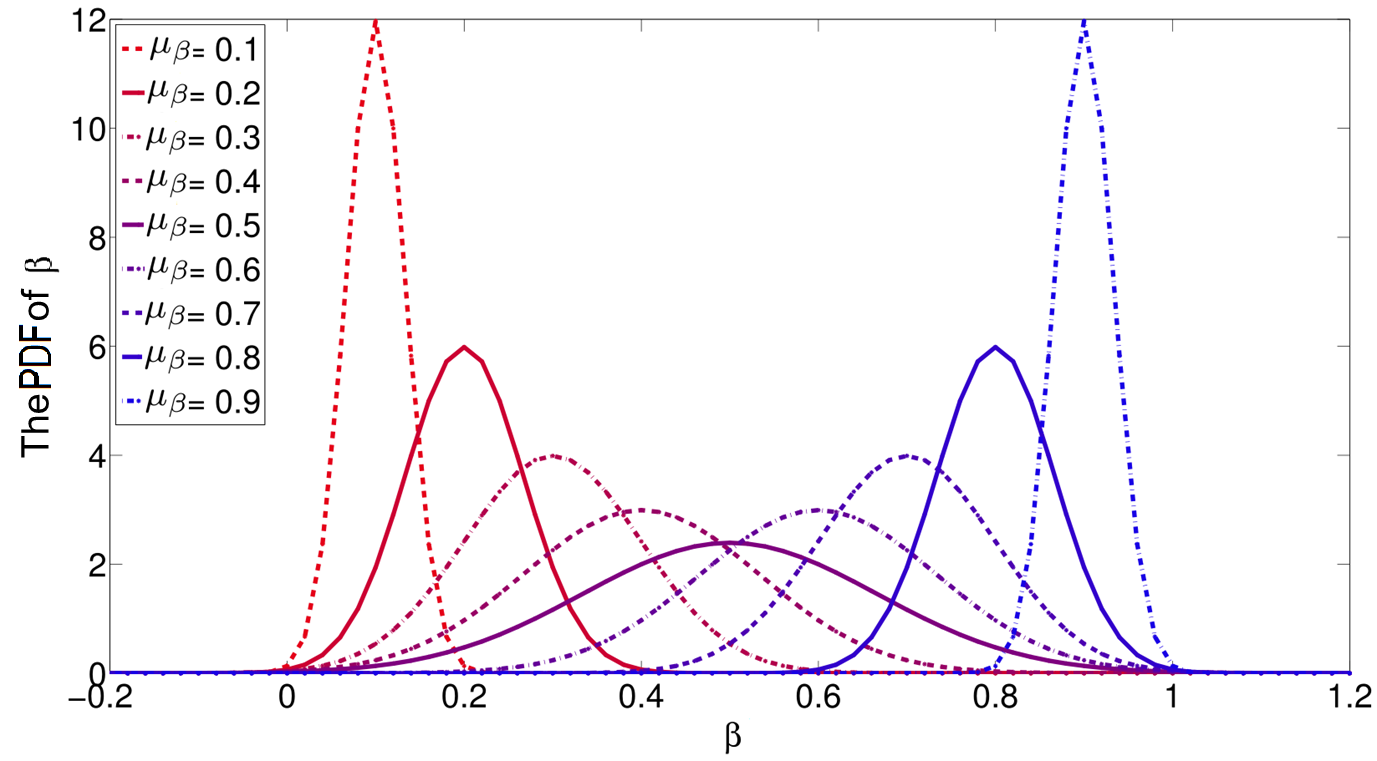}
\caption{$\beta$ is distributed with normal PDF with a mean value of $\mu_{\beta}$.}
\label{fig:betas}
\end{figure}

The $\sigma_{_{\beta}}$ value should also be selected in such a way that the probability of $\beta$ falling outside $\left [0, 1 \right]$ should be negligible. In this paper we set $\sigma_{_{\beta}}$ to be
\begin{equation}
\sigma_{_{\beta}}=\frac{0.5-\left | \mu_{\beta} - 0.5 \right |}{3},
\end{equation}
thus the probability of $\beta$ falling outside $\left [0, \; 1 \right]$ will be 0.1\% as shown in Fig. \ref{fig:betas}.

Figure \ref{fig:cdfs}(b) shows the CDF of the UE density of network after moving UEs towards SAs with different $\beta$ values using the enhanced method.

\begin{figure*}[!t]
\centering
\subfloat[$\alpha$ = 0 and $\beta$ = 0]{\includegraphics[width=0.23\textwidth]{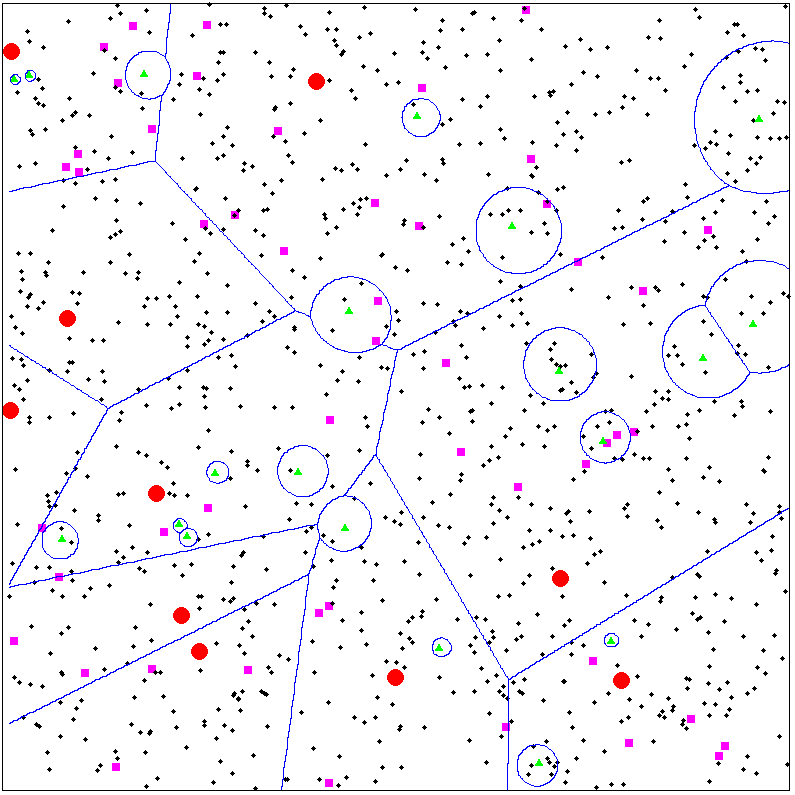}}
~
\subfloat[$\alpha$ = 0 and $\beta$ = 0.3]{\includegraphics[width=0.23\textwidth]{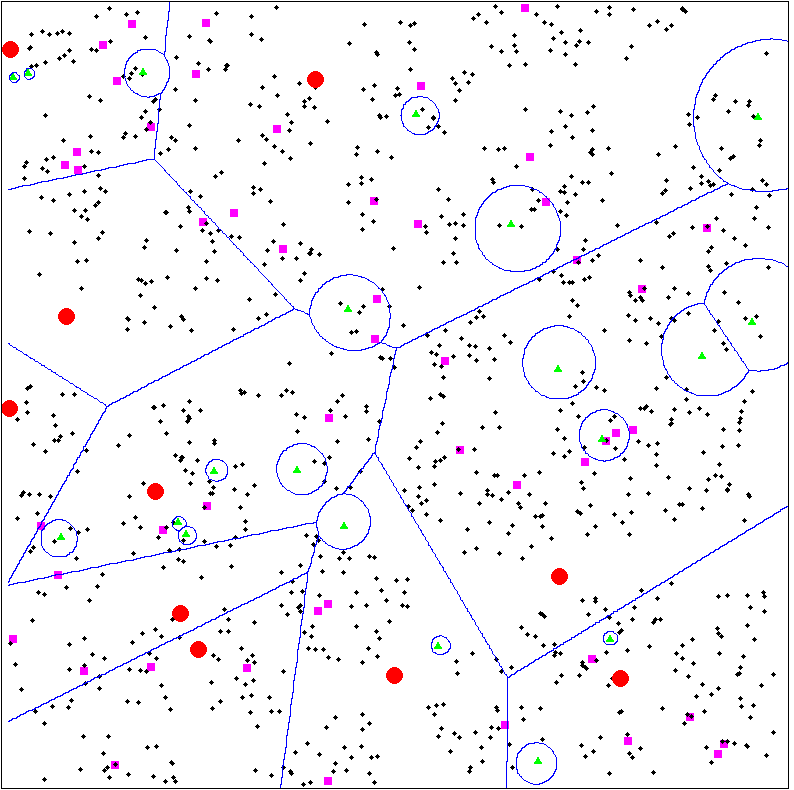}}
~
\subfloat[$\alpha$ = 0 and $\beta$ = 0.6]{\includegraphics[width=0.23\textwidth]{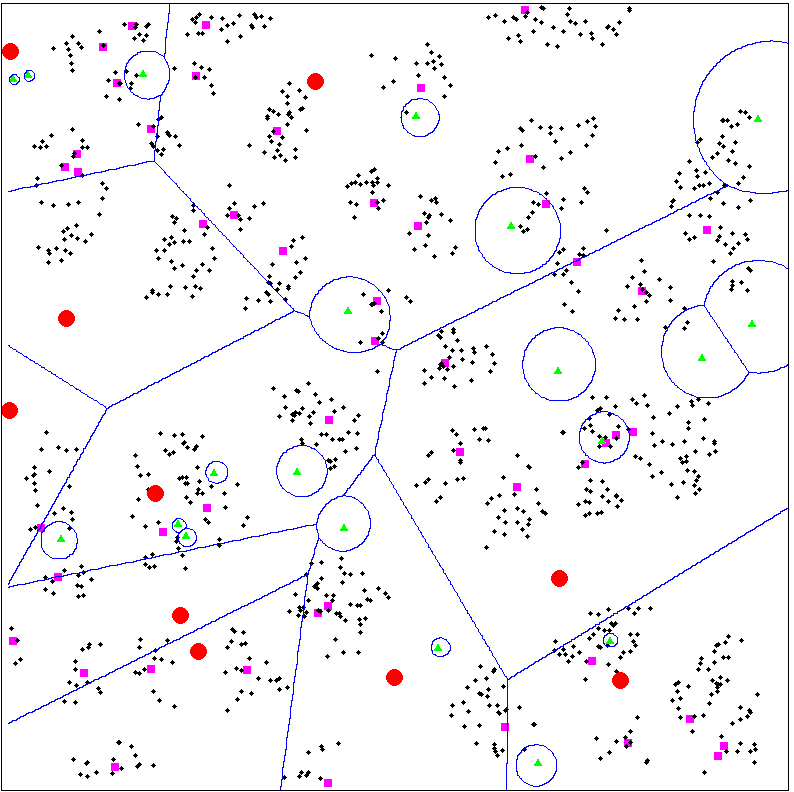}}
~
\subfloat[$\alpha$ = 0 and $\beta$ = 0.9]{\includegraphics[width=0.23\textwidth]{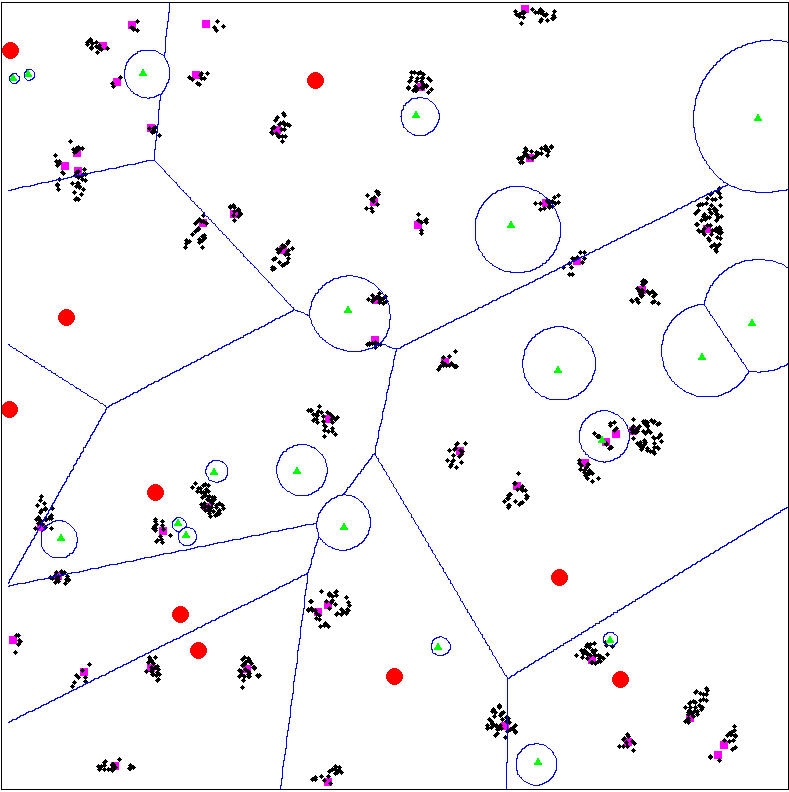}}

\subfloat[$\alpha$ = 0.3 and $\beta$ = 0]{\includegraphics[width=0.23\textwidth]{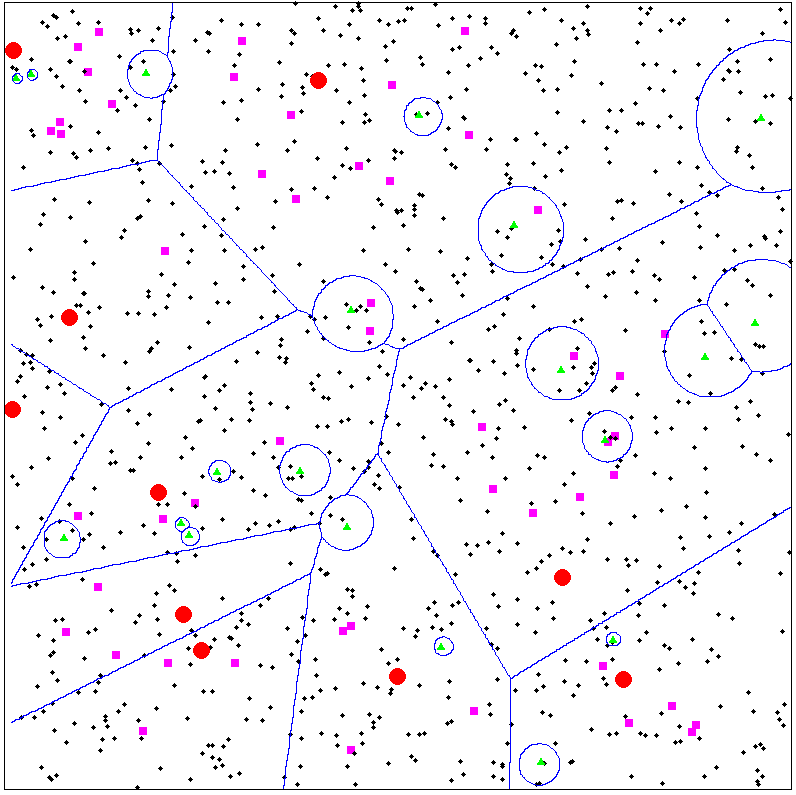}}
~
\subfloat[$\alpha$ = 0.3 and $\beta$ = 0.3]{\includegraphics[width=0.23\textwidth]{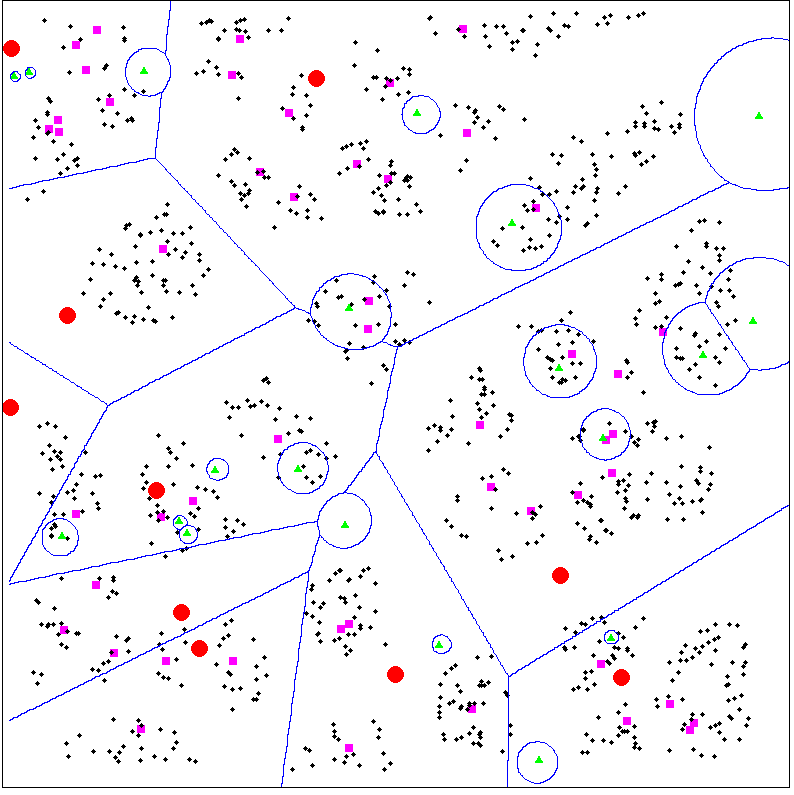}}
~
\subfloat[$\alpha$ = 0.3 and $\beta$ = 0.6]{\includegraphics[width=0.23\textwidth]{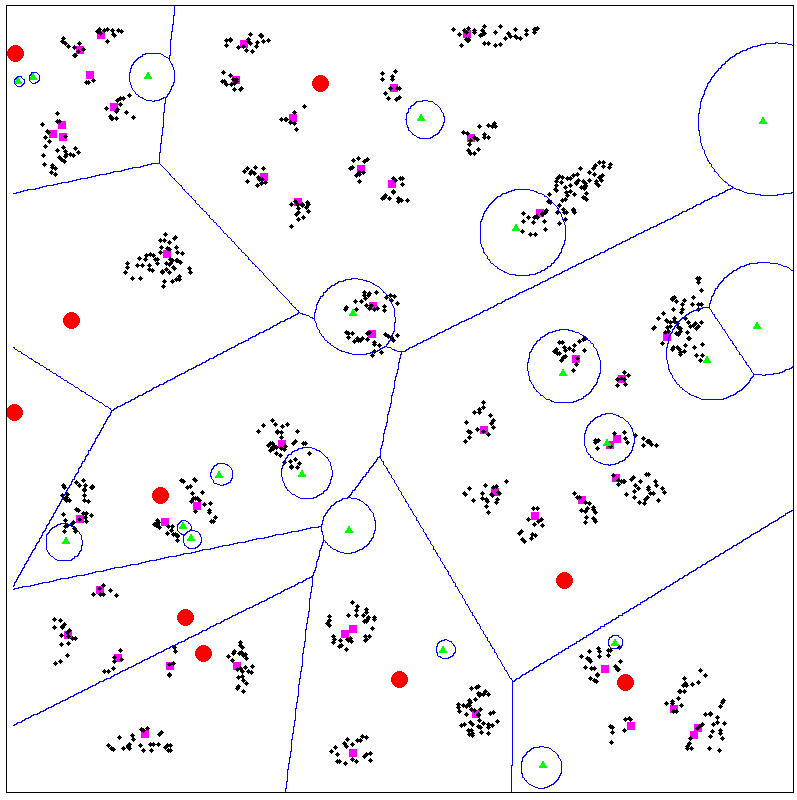}}
~
\subfloat[$\alpha$ = 0.3 and $\beta$ = 0.9]{\includegraphics[width=0.23\textwidth]{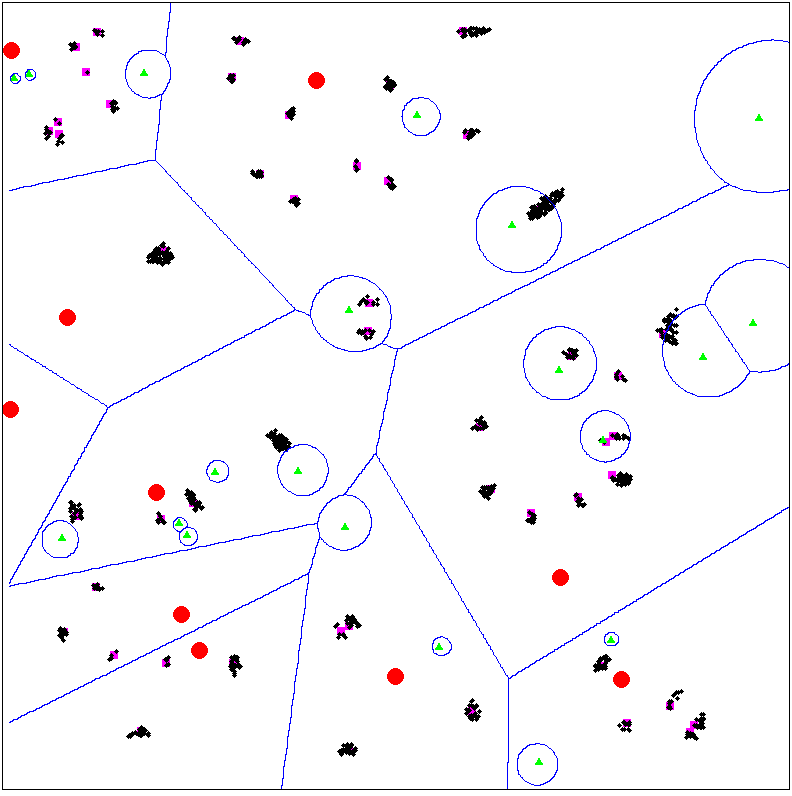}}

\subfloat[$\alpha$ = 0.6 and $\beta$ = 0]{\includegraphics[width=0.23\textwidth]{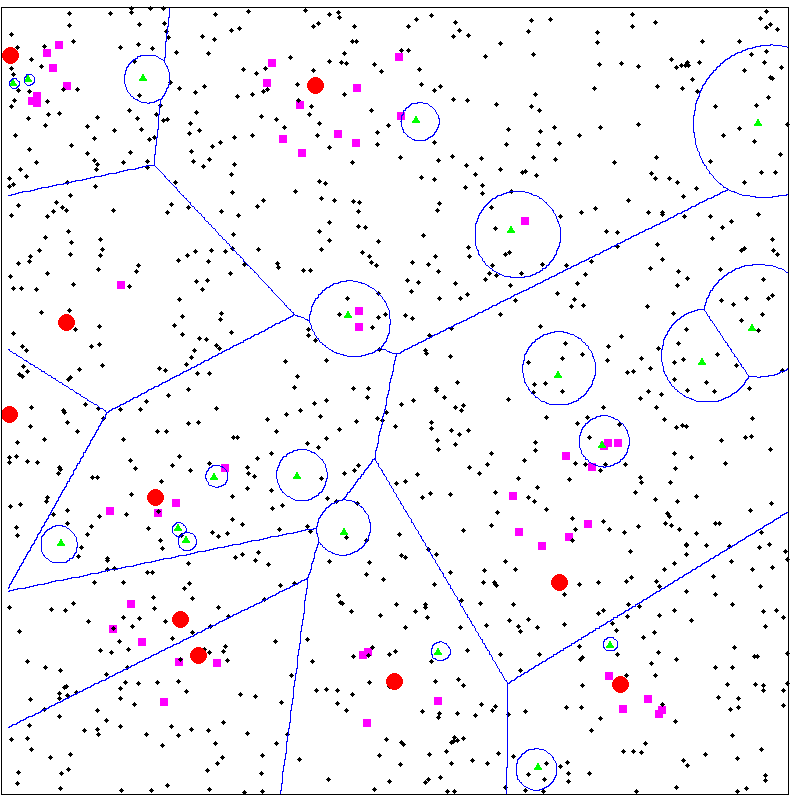}}
~
\subfloat[$\alpha$ = 0.6 and $\beta$ = 0.3]{\includegraphics[width=0.23\textwidth]{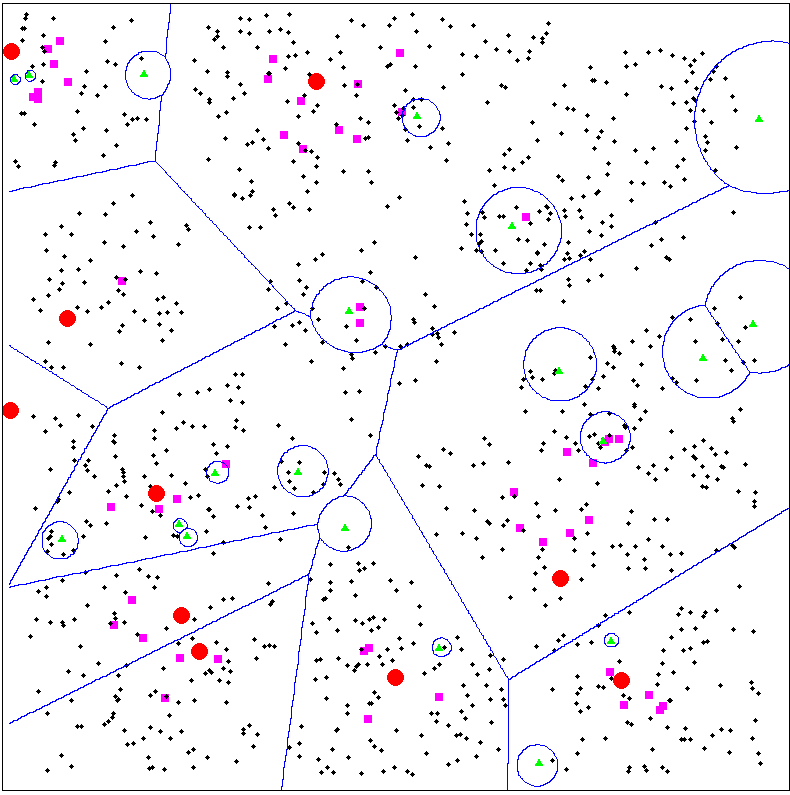}}
~
\subfloat[$\alpha$ = 0.6 and $\beta$ = 0.6]{\includegraphics[width=0.23\textwidth]{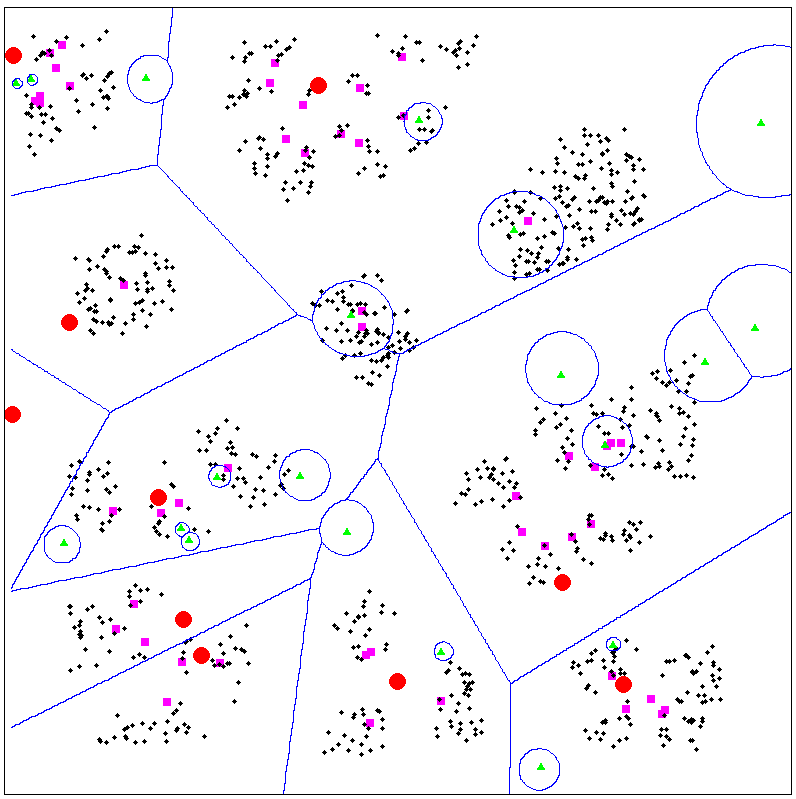}}
~
\subfloat[$\alpha$ = 0.6 and $\beta$ = 0.9]{\includegraphics[width=0.23\textwidth]{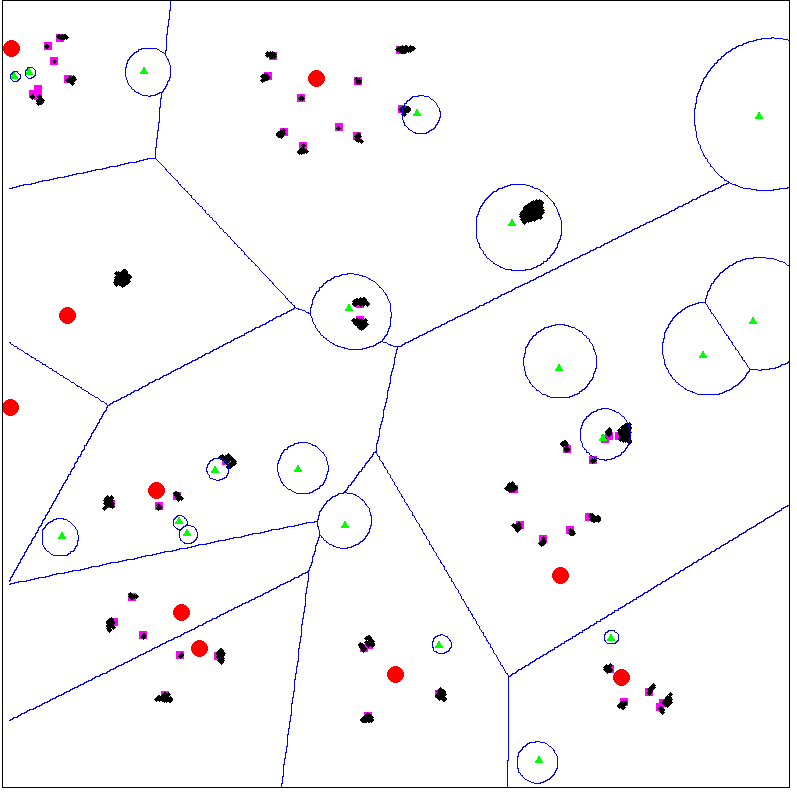}}

\subfloat[$\alpha$ = 0.9 and $\beta$ = 0]{\includegraphics[width=0.23\textwidth]{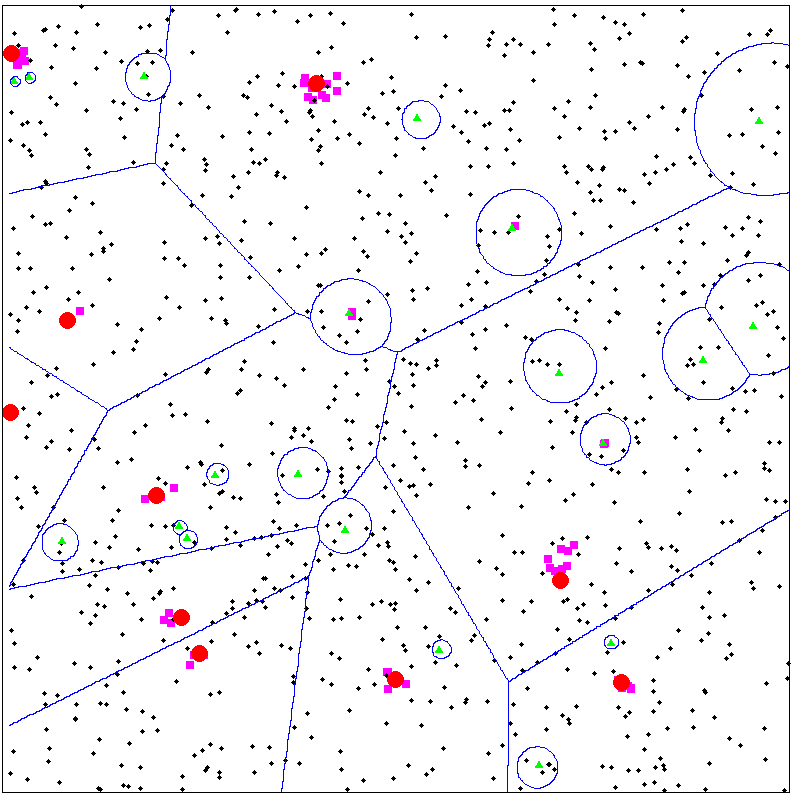}}
~
\subfloat[$\alpha$ = 0.9 and $\beta$ = 0.3]{\includegraphics[width=0.23\textwidth]{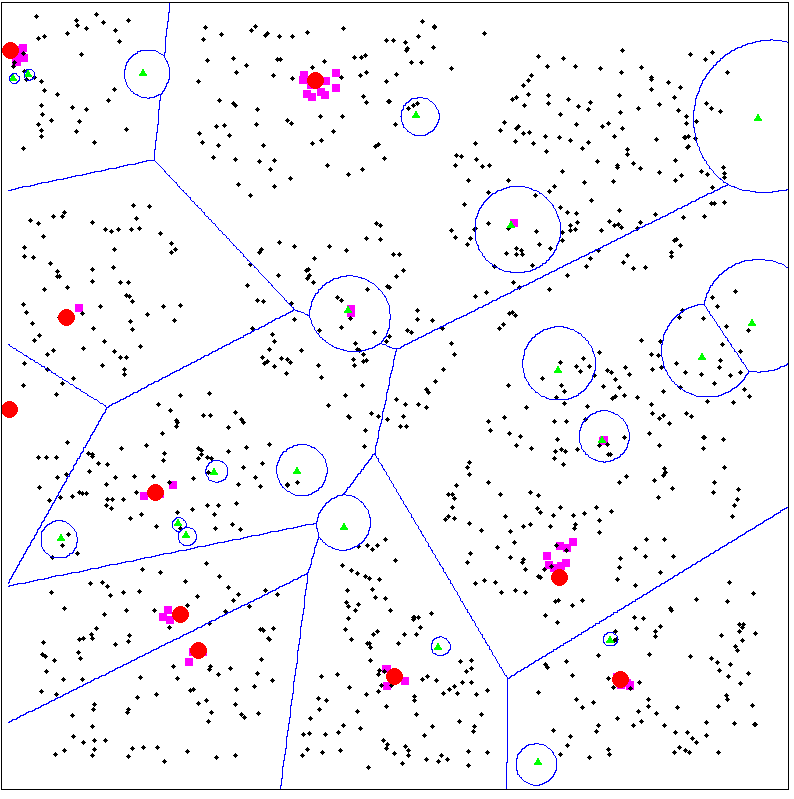}}
~
\subfloat[$\alpha$ = 0.9 and $\beta$ = 0.6]{\includegraphics[width=0.23\textwidth]{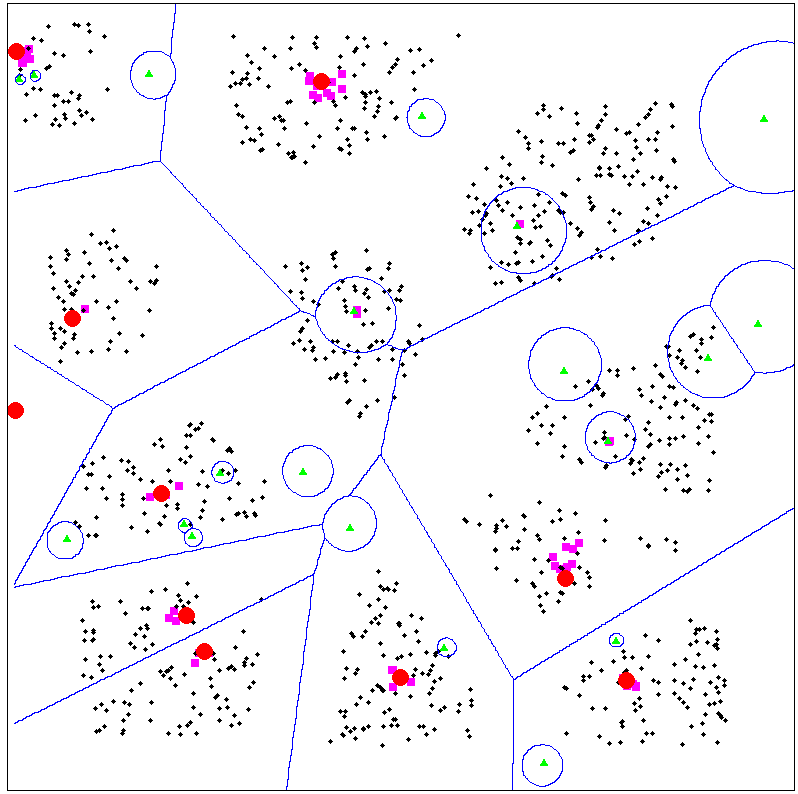}}
~
\subfloat[$\alpha$ = 0.9 and $\beta$ = 0.9]{\includegraphics[width=0.23\textwidth]{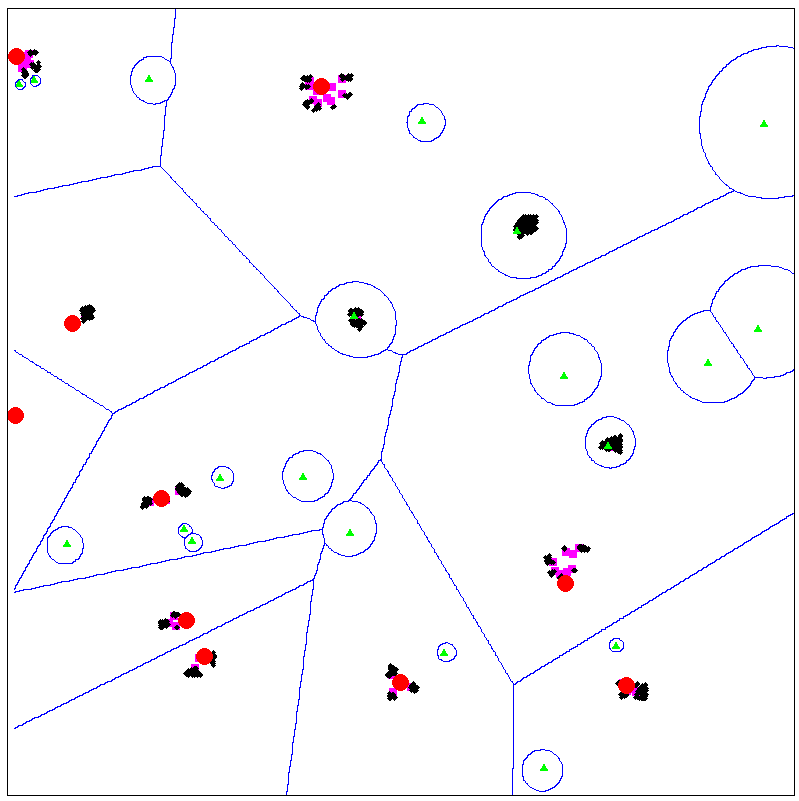}}
\caption{Traffic distribution scenarios: By regulating only two parameters, $\alpha$ (the closeness indicator of social attractors to BSs), and $\beta$ (the closeness indicator of UEs to social attractors) from 0 to 1, traffic distributions with various properties (CoV and BS-correlation) can be generated. The big circles denote macro-BSs, the triangles denote pico-BSs, the lines are the edges of the weighted-Voronoi tessellation of BSs, squares denote social attractors and small circles represent UEs. The range of the area is from 0 to 1000 meters in each dimension.}
\label{fig:scenarios}
\end{figure*}

Figure \ref{fig:scenarios} illustrates various scenarios with different characteristics which can be generated by the proposed traffic generation method.

In our modeling methodology, the TGIPs are $\alpha$ and $\beta$ which are the internal parameters of the traffic generator, and the traffic statistics of interest are $C$ and $\rho$. To make the mapping (i.e., the look-up table generation) from the statistics ($C$ and $\rho$ in our method) to the TGIPs ($\alpha$ and $\beta$ in our method), we first generate traffic patterns with different values of TGIPs and then measure the associated statistics. Using a fitting procedure, the resulting mapping can be expressed as
\begin{equation}
C = F_{1}(\alpha,\beta),
\end{equation}
and
\begin{equation}
\rho = F_{2}(\alpha,\beta).
\end{equation}

$F_{1}$ and $F_{2}$ are monotonically non-decreasing functions due to the fact that the CoV and correlation coefficient values are both non-decreasing with increasing $\alpha$ and $\beta$. Therefore, the final maps from the desired statistics to TGIPs can be obtained by the corresponding inverse operations as follows:
\begin{equation}
\alpha = H_{1}(C,\rho),
\end{equation}
and
\begin{equation}
\beta = H_{2}(C,\rho).
\end{equation}


\section{Numerical Results}
\label{sec:results}

Following the methodology described in Section \ref{sec:methodology}, the traffic measurement method discussed in Section \ref{sec:measurement}, and the traffic generation model described in Section \ref{sec:generation}, the simulation results are presented in this section. The traffic modeling results are presented in Section \ref{subsec:modelingresults}, and the heterogeneous wireless cellular network performance analysis results are presented in Section \ref{subsec:performanceresults}.

\subsection{Traffic Modeling Results}
\label{subsec:modelingresults}

Considering a two-tier superposition of BSs (i.e., macro-BSs and pico-BSs), 10 macro-BSs and 20 pico-BSs are distributed in a 1000 m $\times$ 1000 m square field. The distribution of BSs is assumed to be PPP. 50 SAs are distributed by an independent PPP. UEs are distributed using the traffic generation method described in Section \ref{sec:generation} with parameters $\alpha$ and $\beta$. The simulation is repeated for 1000 random drops.

\begin{figure}
\centering
\includegraphics[width=\columnwidth]{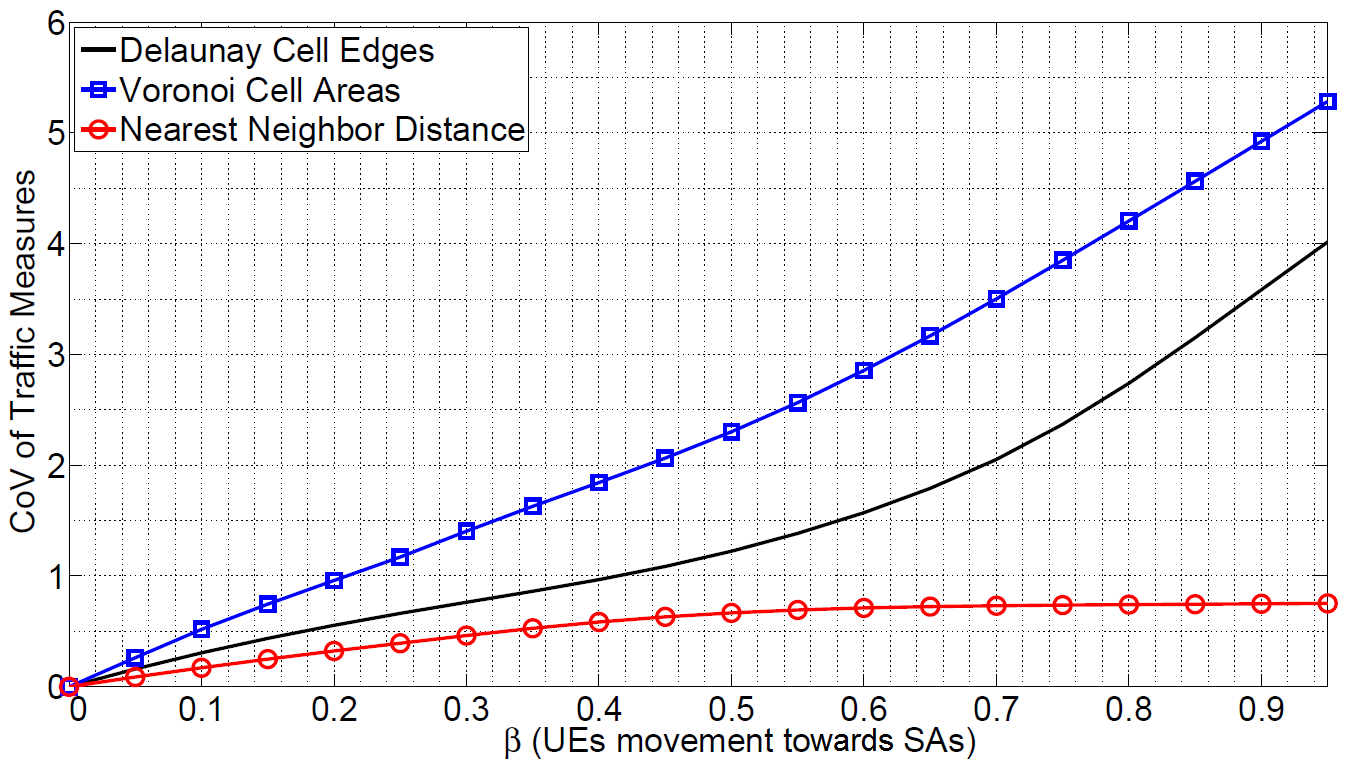}
\caption{To compare the proposed distance-based traffic measures with the existing nearest-neighbor distance measure, we fixed the $\alpha$ value to zero ($\alpha=0$) and changed the $\beta$ value from 0 to 1 and calculated the CoV of different measures. Nearest-neighbor distance cannot capture the traffic heterogeneity and stays constant in super-Poisson region.}
\label{fig:measures}
\end{figure}

The first step in traffic measurement is to select the appropriate traffic measure. To compare the proposed spatial traffic measures with the existing distance-based nearest-neighbor distance measure, with $\alpha$ fixed at 0, we increased $\beta$ from 0 to 1 and calculated different traffic measures for each $\beta$ value. Figure \ref{fig:measures} illustrates the CoV values of various distance-based traffic measures introduced in this paper.

\begin{figure}
\centering
\includegraphics[width=\columnwidth]{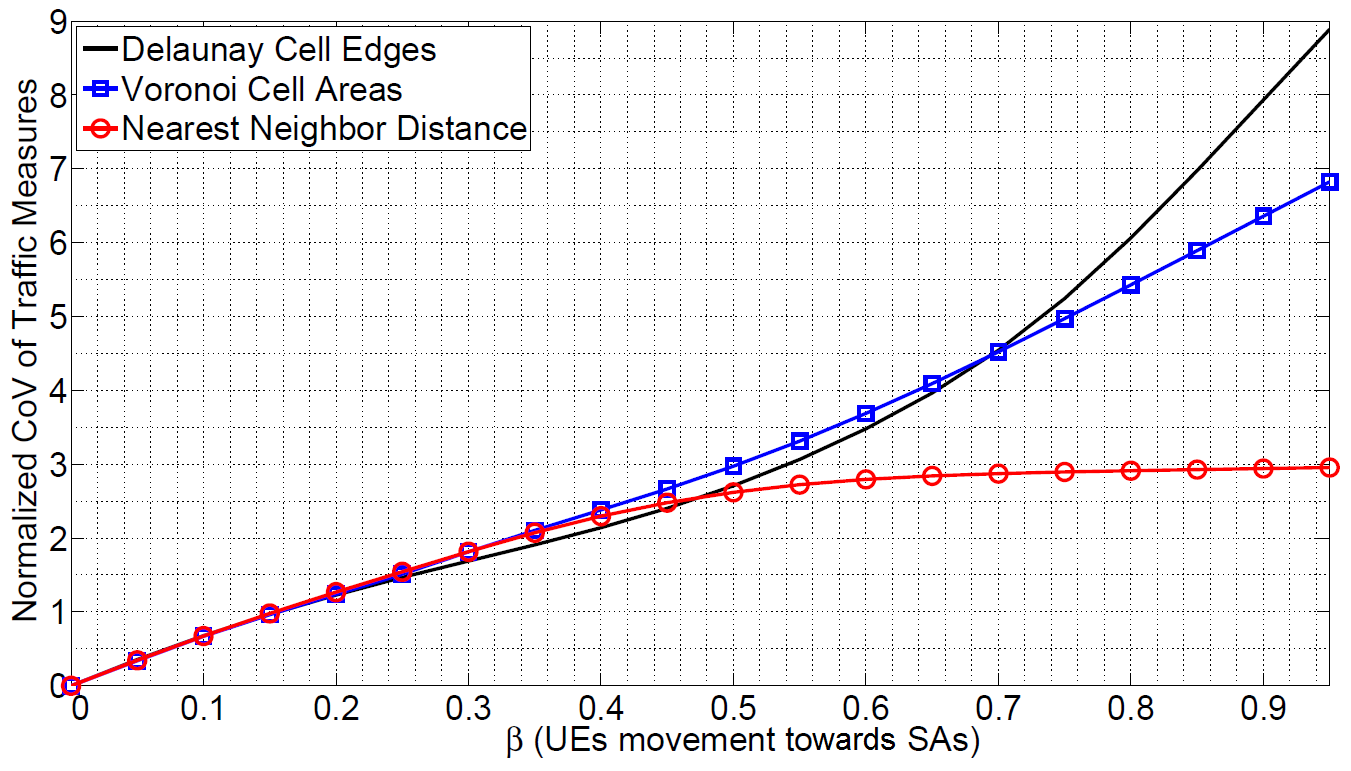}
\caption{To be consistent with time domain traffic measurement, we divide the CoV of the space traffic measures by the convergence value at Poisson patterns. The normalized CoV values are used in this paper.}
\label{fig:measuresnorm}
\end{figure}

The CoVs of all the measures are normalized to 1 at the Poisson case by dividing the CoVs by their expected values (presented in Table \ref{tab:distmeasures}). Figure \ref{fig:measuresnorm} shows the normalized results.

As it can be seen in Fig. \ref{fig:measuresnorm}, the nearest-neighbor distance measure cannot capture the traffic heterogeneity as it stays constant in the super-Poisson region. On the other hand, the two proposed measures capture the traffic heterogeneity for all CoV values. The Voronoi cell area is preferable to the Delaunay cell edge length for two reasons: First, there exists a Voronoi cell area associated with each user, while a Delaunay cell edge length cannot be associated with one particular UE. Secondly, the slope of the Voronoi cell area with respect to $\beta$ is higher than the slope of the Delaunay cell edge length with respect to $\beta$; thus the Voronoi cell area can capture the traffic heterogeneity with a higher resolution. In the remainder of this paper we use the Voronoi cell area as the traffic measure.

\begin{figure}
\centering
\includegraphics[width=\columnwidth]{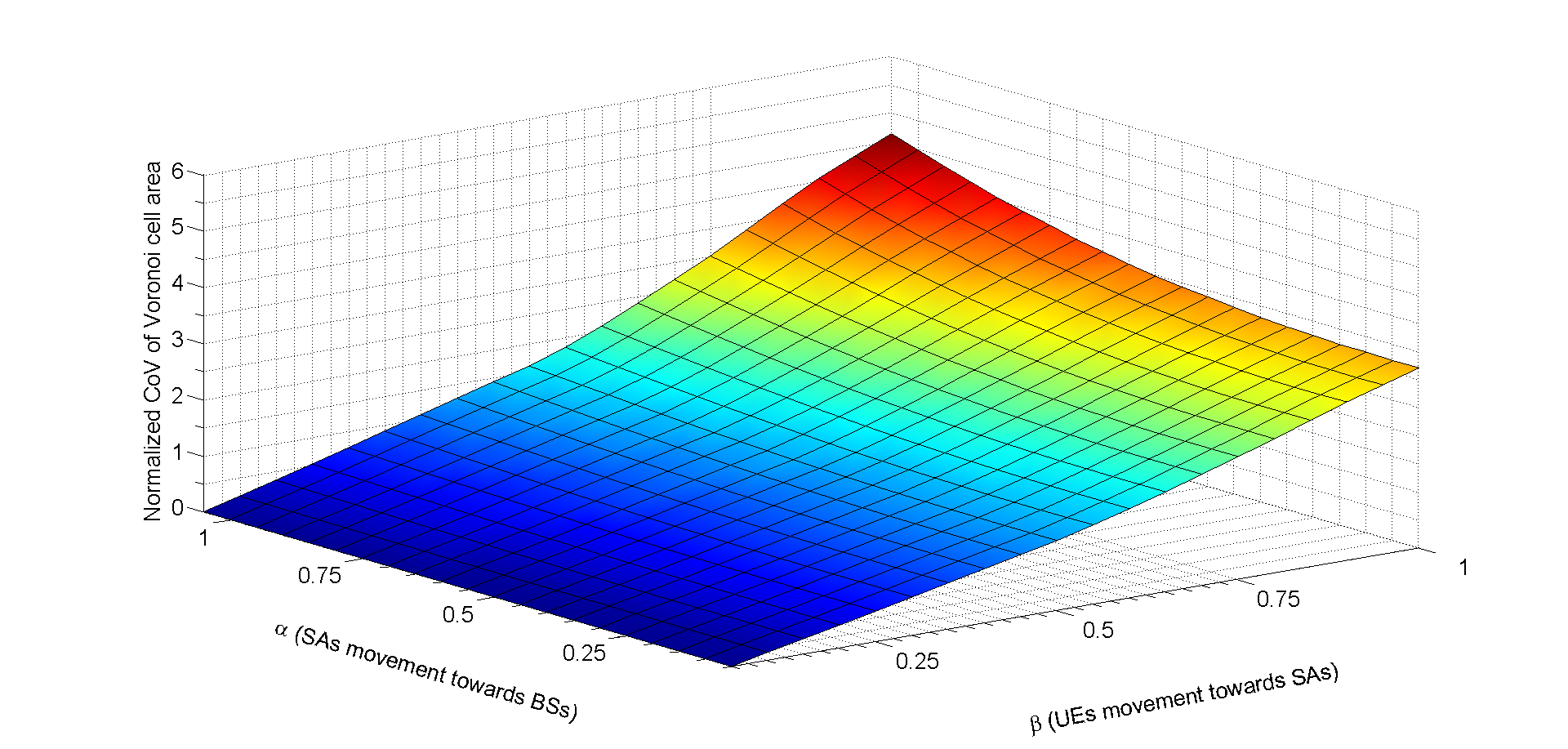}
\caption{The normalized CoV of Voronoi cell area is calculated for traffic generated with different values of $\alpha$ and $\beta$.}
\label{fig:cv_map}
\end{figure}

\begin{figure}
\centering
\includegraphics[width=\columnwidth]{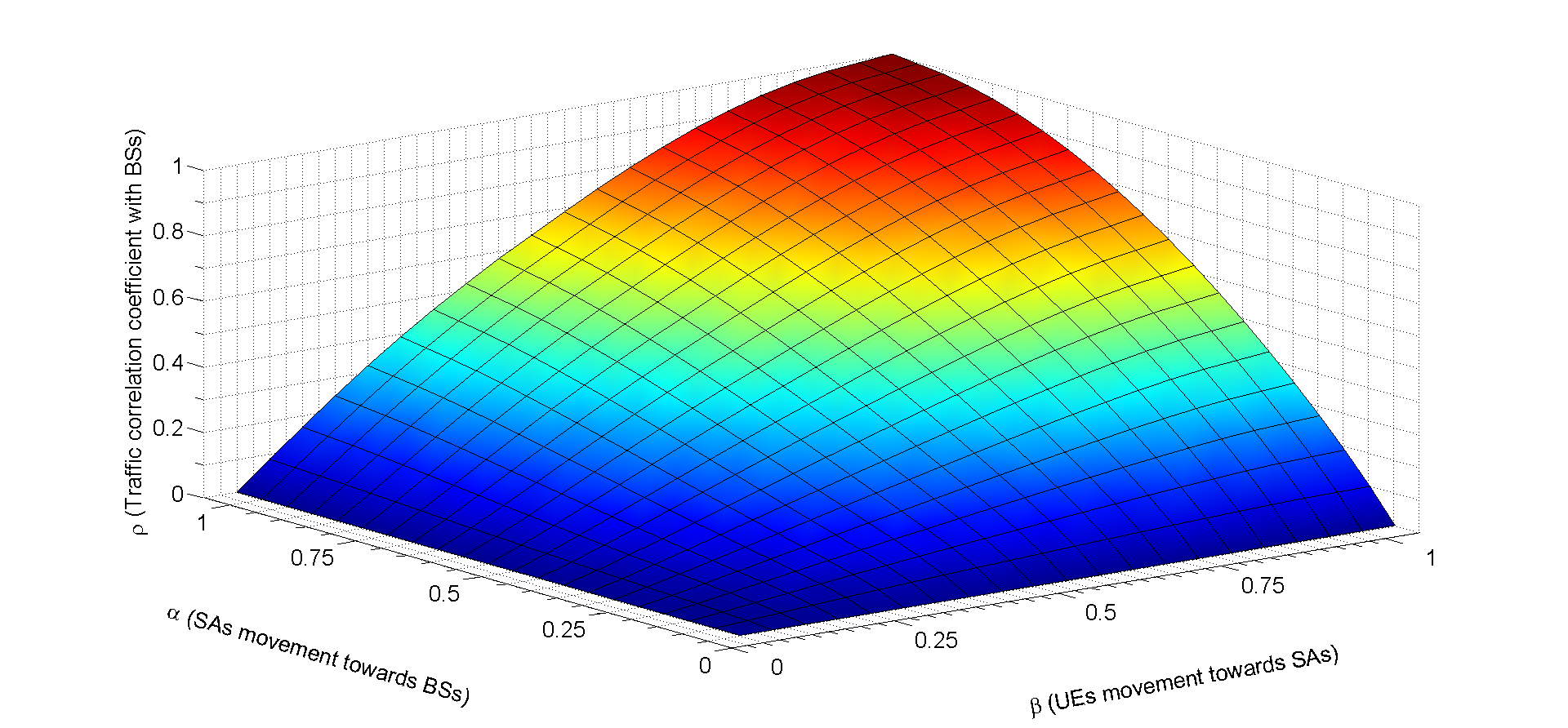}
\caption{The correlation coefficient (refer to (\ref{eq:rho})) between UEs and BSs is calculated for traffic generated with different values of $\alpha$ and $\beta$.}
\label{fig:xc_map}
\end{figure}

The next step in traffic modeling is to generate a map from TGIPs to traffic statistics. Figures \ref{fig:cv_map} and \ref{fig:xc_map} demonstrate the calculated CoV and correlation coefficient values, respectively, for different values of $\alpha$ and $\beta$, both ranging from 0 to 1.

An important observation from the maps in Figures \ref{fig:cv_map} and \ref{fig:xc_map} is that the CoV and BS-correlation are not independent. In other words, the range of achievable CoV values for each correlation coefficient value is different. This is due to the fact that when UEs are biased towards BSs, UEs are attracted towards the points of interest, and this automatically shapes clusters; which results in a degree of clustering and increases the CoV.

\begin{figure}
\centering
\includegraphics[width=\columnwidth]{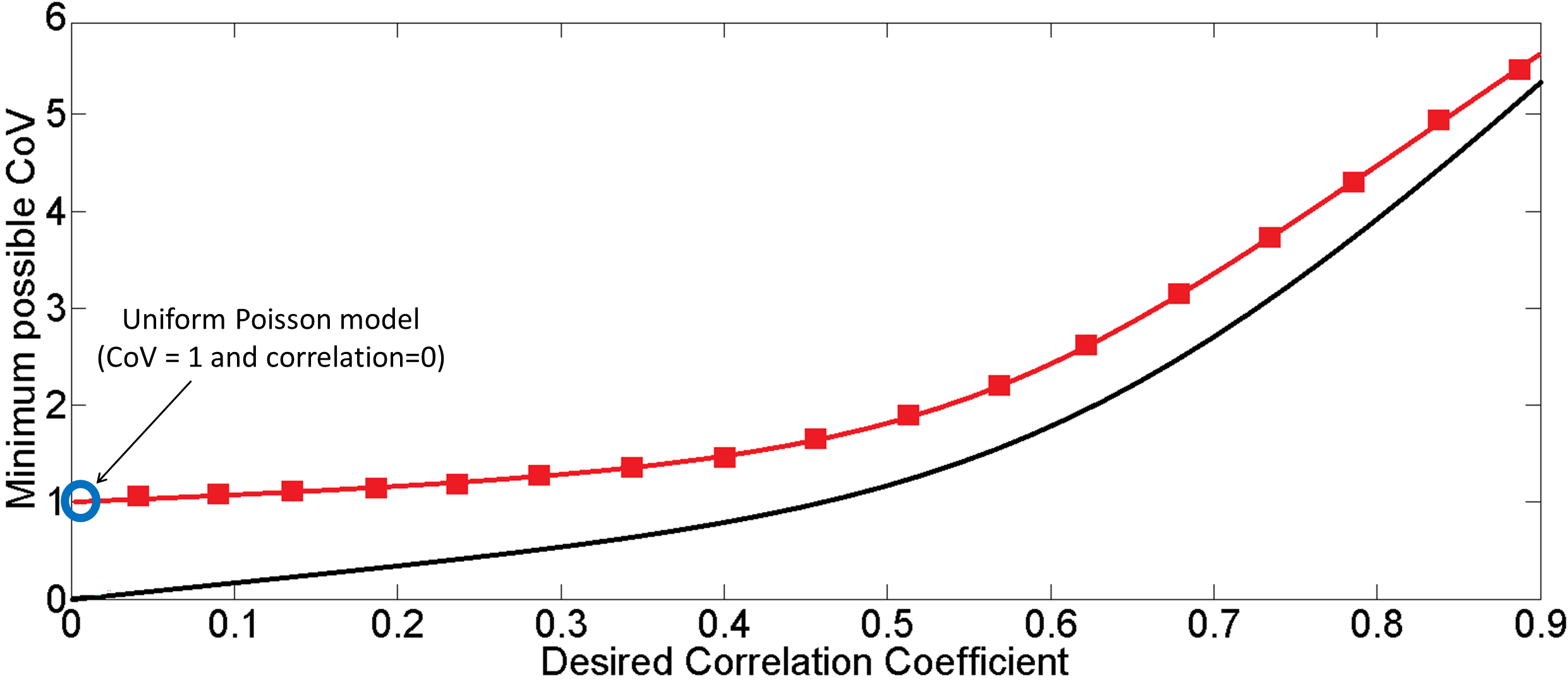}
\caption{The feasible normalized CoV values for different values of correlation coefficient are shown. Traffic with high correlation with BSs cannot have low normalized CoV values because high correlation means that UEs are gathered at cell centers. All the region above the solid line is the possible traffic generation using our proposed method. The square-marked line shows the possible traffic generation using the model presented in \cite{dhillon2012modeling1}.}
\label{fig:feasibleregion}
\end{figure}

To make a comparison with the existing models in the literature, we should note that the prevailing spatial model used for UE distribution in the literature is the uniform PPP. To compare the PPP model to our model we should say that the PPP corresponds to just one specific point in the whole space of the possible spatial distribution situations (CoV (heterogeneity) =1 and $\rho$ (BS-correlation) =0). There are few models in the literature which consider heterogeneous or BS-correlated traffic modeling. To the best of the authors' knowledge the most appropriate of those is the one presented in \cite{dhillon2012modeling1} which is a BS-correlated model. However, the model in \cite{dhillon2012modeling1} is capable of generating a limited sub-space of the whole traffic possibilities. Figure \ref{fig:feasibleregion} illustrates the feasible region of CoV and correlation coefficient values generated by our model versus the model presented in \cite{dhillon2012modeling1}.

\begin{figure}
\centering
\includegraphics[width=\columnwidth]{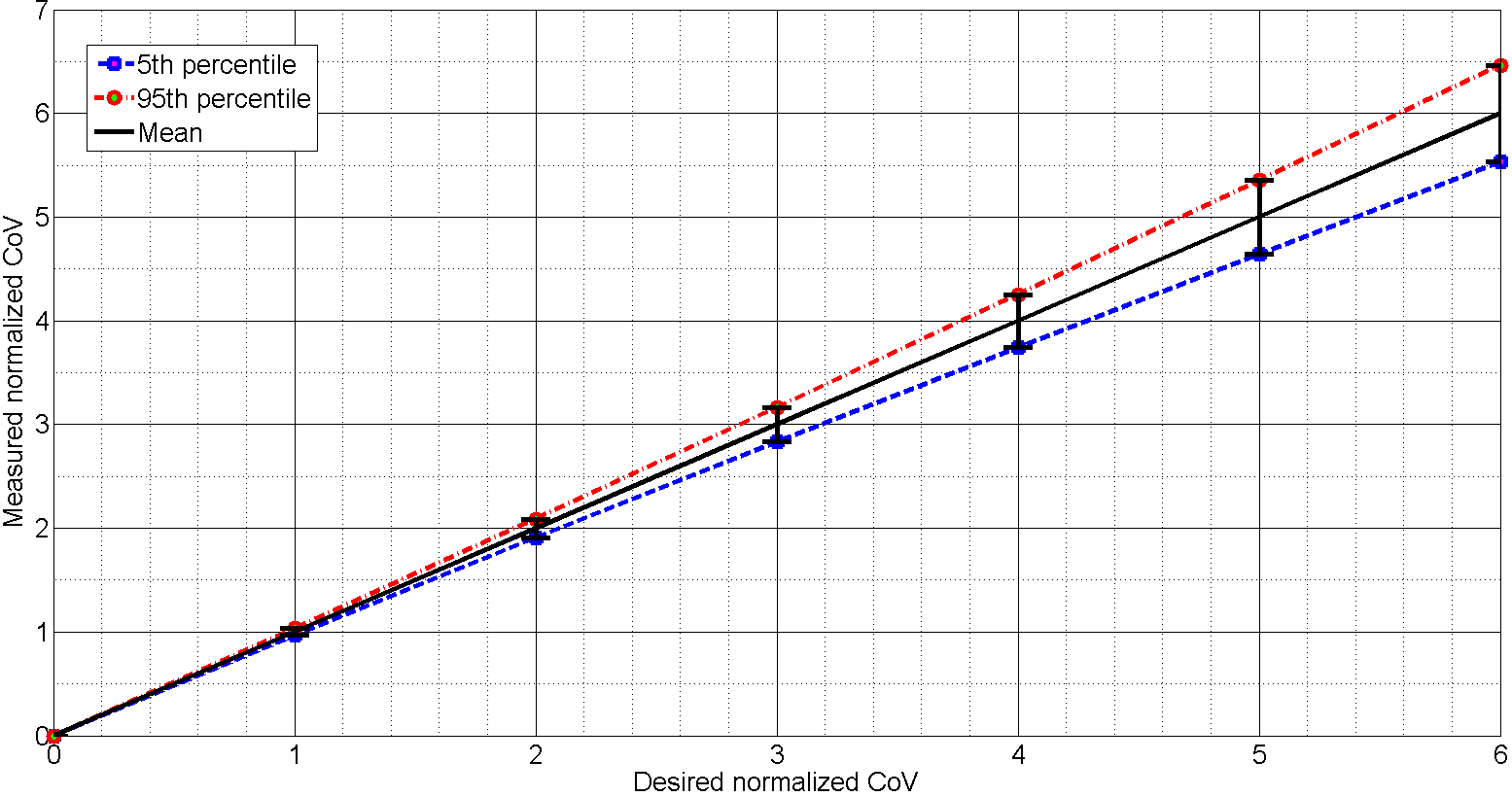}
\caption{The horizontal axis shows the desired normalized CoV. The desired normalized CoV is mapped to the associated $\alpha$ and $\beta$ values which are used to generate traffic. The vertical axis shows the calculated normalized CoV from the generated traffic patterns (for each desired $C$ value, the traffic is generated for all feasible $\rho$ values and the CoV is averaged out).}
\label{fig:inoutcv}
\end{figure}

\begin{figure}
\centering
\includegraphics[width=\columnwidth]{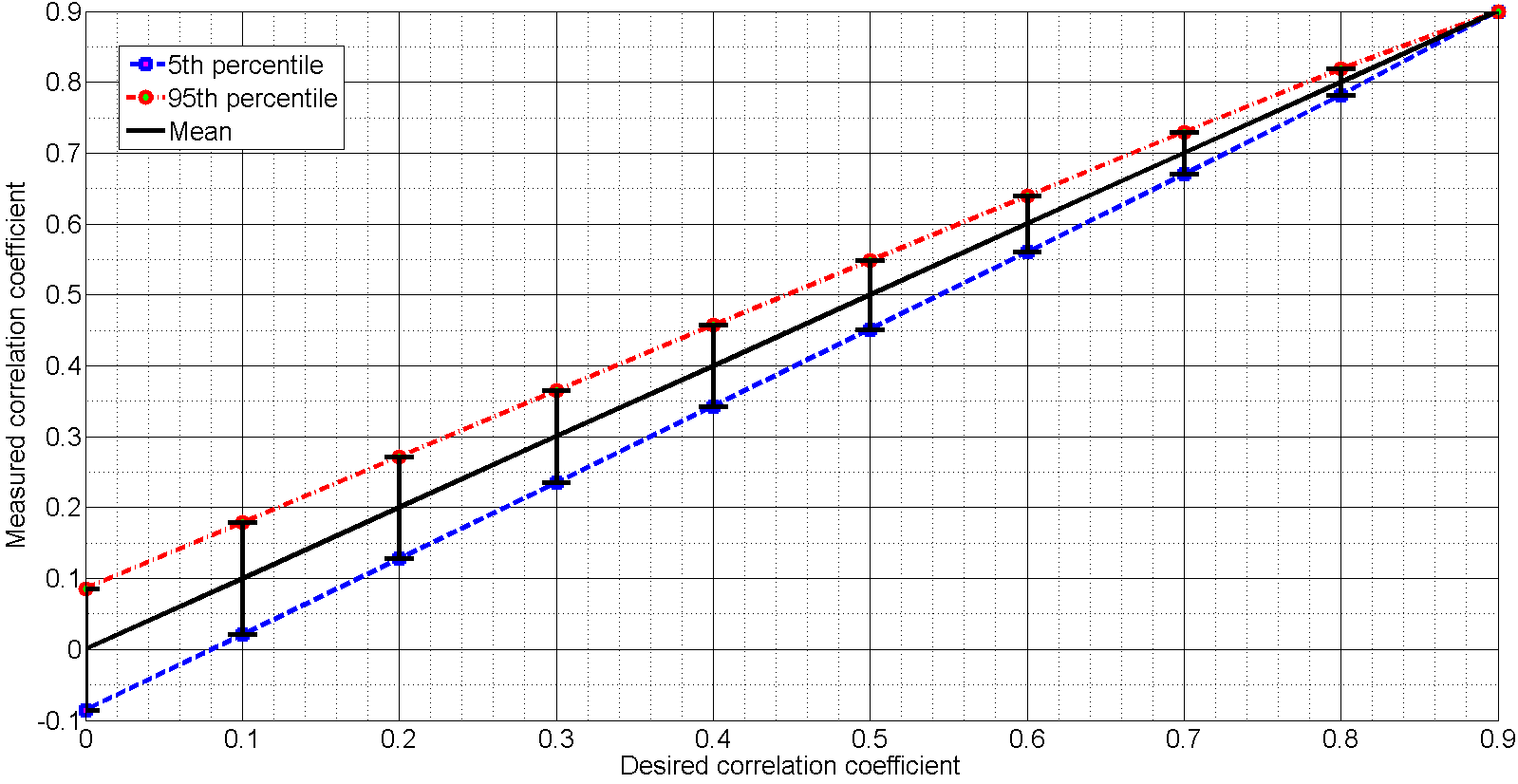}
\caption{The horizontal axis shows the desired $\rho$. The desired $\rho$ is mapped to the associated $\alpha$ and $\beta$ values which are used to generate traffic. The vertical axis shows the calculated $\rho$ from the generated traffic patterns (for each desired $\rho$ value, the traffic is generated for all feasible normalized CoV values and the correlation coefficient is averaged out).}
\label{fig:inoutxc}
\end{figure}

The next step in traffic modeling is to translate the desired statistics to appropriate TGIPs which can generate traffic with the desired statistics. Towards that end, the statistical properties of the generated traffic are measured. Since the process of traffic generation is a random process (e.g., $\beta$ is a random variable), the measured statistics are not constant and have a deviation around their expected values. The deviation of the measured statistics from the desired statistics determines the accuracy of the method. Figures \ref{fig:inoutcv} and \ref{fig:inoutxc} illustrate the measured statistics versus the desired statistics.

\subsection{Network Performance Analysis Results}
\label{subsec:performanceresults}

The performance of a downlink LTE cellular network is analyzed in this section. In a field of 1000 m $\times$ 1000 m, UEs are distributed using the traffic generation method described in Section \ref{sec:generation}. The simulation parameters are summarized in Table \ref{tab:parameters} \cite{ICT10WINNER,DBNRS_PIMRC10}.

\begin{table}
\centering
\caption{\normalfont{Simulation parameters.}}
\label{tab:parameters}
    \begin{tabular}{|m{4cm}|m{4cm}|}
    \hline
    \textbf{PARAMETER}                   & \textbf{VALUE}                                             \\ \hline
    Mean number of UEs          & 1000                                              \\ \hline
    Number of macro-BSs         & 10                                                \\ \hline
    Number of pico-BSs          & 20                                                \\ \hline
    Number of SAs               & 50                                                \\ \hline
    BSs distribution            & PPP                                     \\ \hline
    BS antenna height           & 10 m                                              \\ \hline
    Number of drops             & 1000                                              \\ \hline
    Bandwidth (downlink)        & 20 MHz                                            \\ \hline
    Noise power per RB          & -174 dBm/Hz                                      \\ \hline
    Carrier frequency           & 2.5 GHz                                           \\ \hline
    Total macro-BS transmit power     & 37 dBm                                   \\ \hline
    Total pico-BS transmit power     & 17 dBm                                   \\ \hline
    Path loss and shaddowing    & Based on UMi scenario \cite{ITU-R2008}            \\ \hline
    BS antenna gain (boresight) & 17 dBi                                            \\ \hline
    UE antenna gain             & 0 dBi                                             \\ \hline
    Time domain traffic model   & Full buffer                                       \\ \hline
    Antenna model               & Omni-directional                                  \\ \hline
    BS down tilt                & 12 degrees                                        \\ \hline
    UE antenna height           & 1.5 m                                             \\ \hline
    Shadowing model             & log-normal with std: LoS: 3, NLoS: 6               \\ \hline
    Fading model                & no fading                                         \\ \hline
    Scheduling                  & Proportional fair \cite{RS_XU_ICC09}              \\ \hline
    \end{tabular}
\end{table}

\begin{figure}
\centering
\includegraphics[width=\columnwidth]{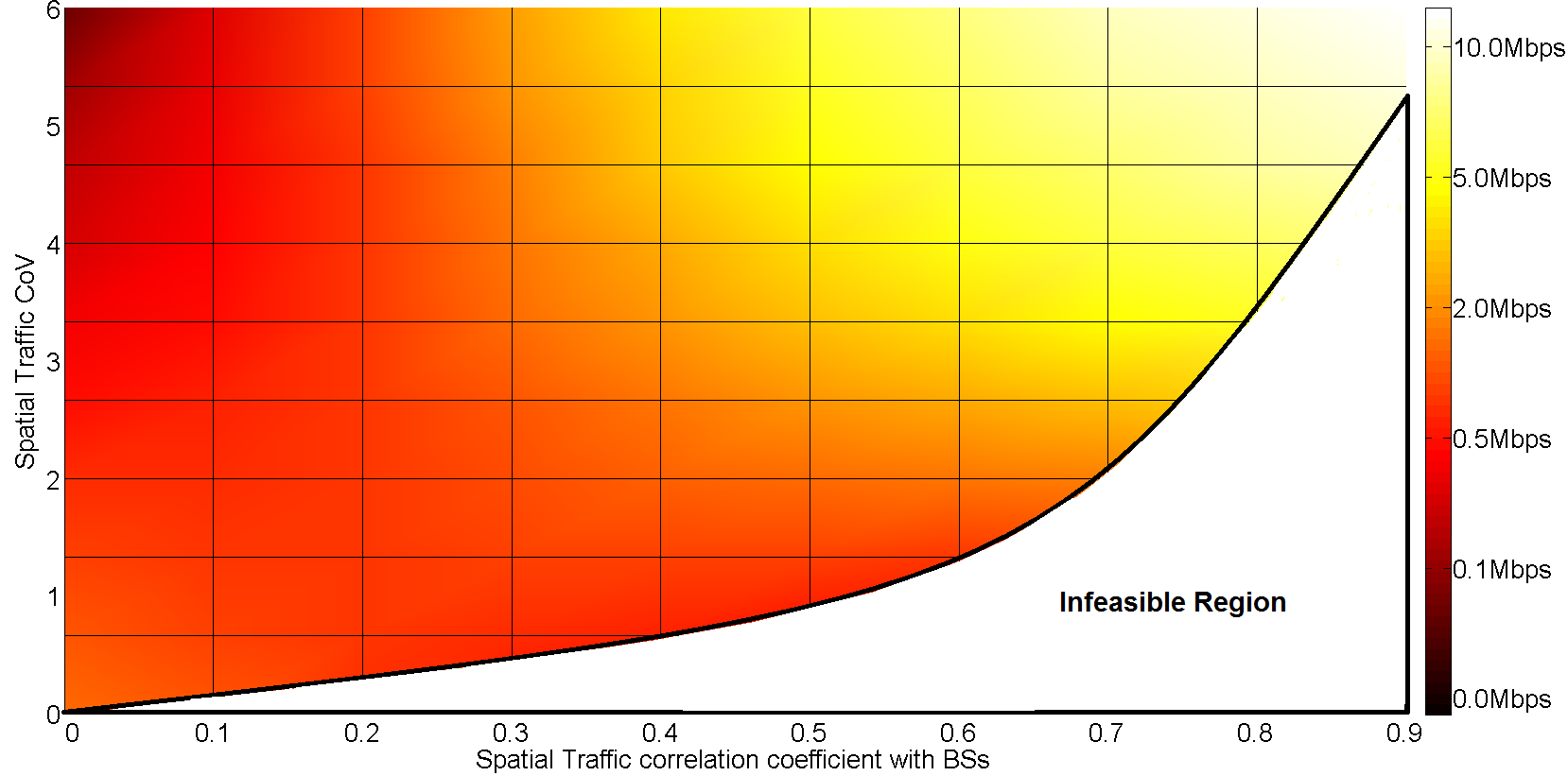}
\caption{The mean user rates is shown for the feasible normalized CoV and correlation coefficient region.}
\label{fig:rates}
\end{figure}

\begin{figure}
\centering
\includegraphics[width=\columnwidth]{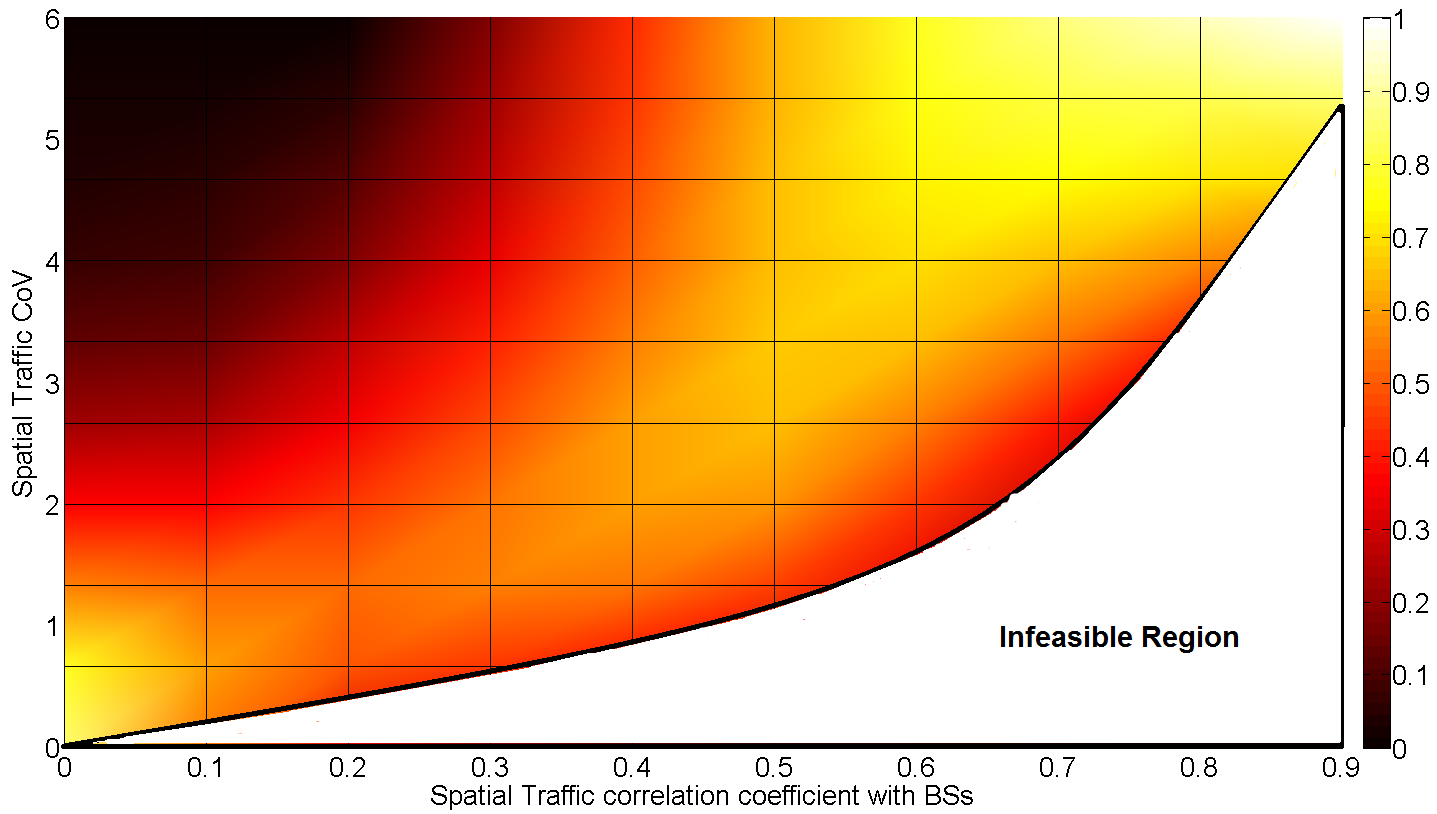}
\caption{The coverage probability with minimum SINR threshold of 10 dB is shown for the feasible normalized CoV and correlation coefficient region.}
\label{fig:coverage}
\end{figure}

The mean value of the rates of all the network users, and the coverage probability with a minimum SINR threshold of 10 dB versus the feasible traffic statistics (for 1000 drops) are presented in Fig. \ref{fig:rates} and Fig. \ref{fig:coverage}, respectively.

The following are the observations made from Figures \ref{fig:rates} and \ref{fig:coverage}:
\begin{itemize}
\item As the correlation coefficient between UE and BS locations increases, the mean user rate and coverage probability also increase as expected, due to the fact that the received signal strength from the serving BS is increased and that from the interfering BSs is decreased. As a result, the SINR statistics are improved.
\item For independent UE and BS locations (or, for low correlation coefficient values), as CoV increases, the mean user rate and coverage probability decrease. This is due to the fact that as CoV increases, the UEs become concentrated in some localities, and as a result, some BSs don't have any UEs connected to them, i.e., the resources of these BSs are wasted. On the other hand, some BSs are overloaded with more than the average number of UEs which have to share the available resources.
\item For correlated UE and BS locations (high correlation coefficient values), as CoV increases, the mean user rate and coverage probability also increase. In these scenarios, with increasing CoV, UEs move more and more towards the BSs (cell centers); as a result, the SINR statistics are improved.
\end{itemize}

\section{Concluding Remarks}
\label{sec:conclusion}

In this paper, a new statistical approach for modeling the spatial traffic in heterogeneous cellular networks is introduced. A number of novel distance-based traffic measures in space are proposed which can be considered as the equivalents of the \textit{iat} in the time domain. Only two statistical parameters are used to regulate the heterogeneity and bias of traffic towards the BSs. Stochastic geometry is used to generate realistic and adjustable traffic. The effects of the realistic traffic modeling on the performance of heterogeneous wireless cellular networks is illustrated.

This work can be extended in a number of directions. It is important to measure the auto-correlation of traffic in space as a fundamental statistic which affects the performance. It is also beneficial to consider other point processes and other Potential functions (refer to (\ref{eq:potential})) for traffic generation. Various techniques, such as the user-in-the-loop (UIL) \cite{RS_APCC10,schoenen2011user,RSComMag2014}, can be used to improve network performance for the heterogeneous and BS-correlated traffic. Finally, a joint time domain and space domain modeling may be developed to obtain a more comprehensive traffic model.

The traffic modeling method presented in this paper is general enough to be applied in other wireless networks as well. For instance, the relaying and routing problems in ad hoc and Wi-Fi networks are closely related, and are dependent on the distribution of the network terminals. In wireless networks, the cell switch-off framework is also highly dependent on the spatial distribution of the traffic.

It is interesting to note that various characteristics of the user distributions in wireless networks can be observed in other phenomena as well. For instance, the similar problems of heterogeneity of distribution or self-similarity arise in micro-cosmic level investigations in chemistry and particle physics and in macro-cosmic level studies in astrophysics.

\section*{Acknowledgment}
\label{sec:ack}

The authors would like to express their gratitude to Dr. Harpreet Dhillon (Virginia Tech, USA) for providing MATLAB code for generating the weighted-Voronoi tessellation, and to Dr. Gamini Senarath and Dr. Ngoc Dao (Huawei Technologies, Canada), for their helpful comments and suggestions.

\appendix[Proof of the Simplest Polynomial Potential Function]
\label{appendix1}

\begin{figure}
\centering
\begin{minipage}{.45\textwidth}
  \centering
  \includegraphics[width=\linewidth]{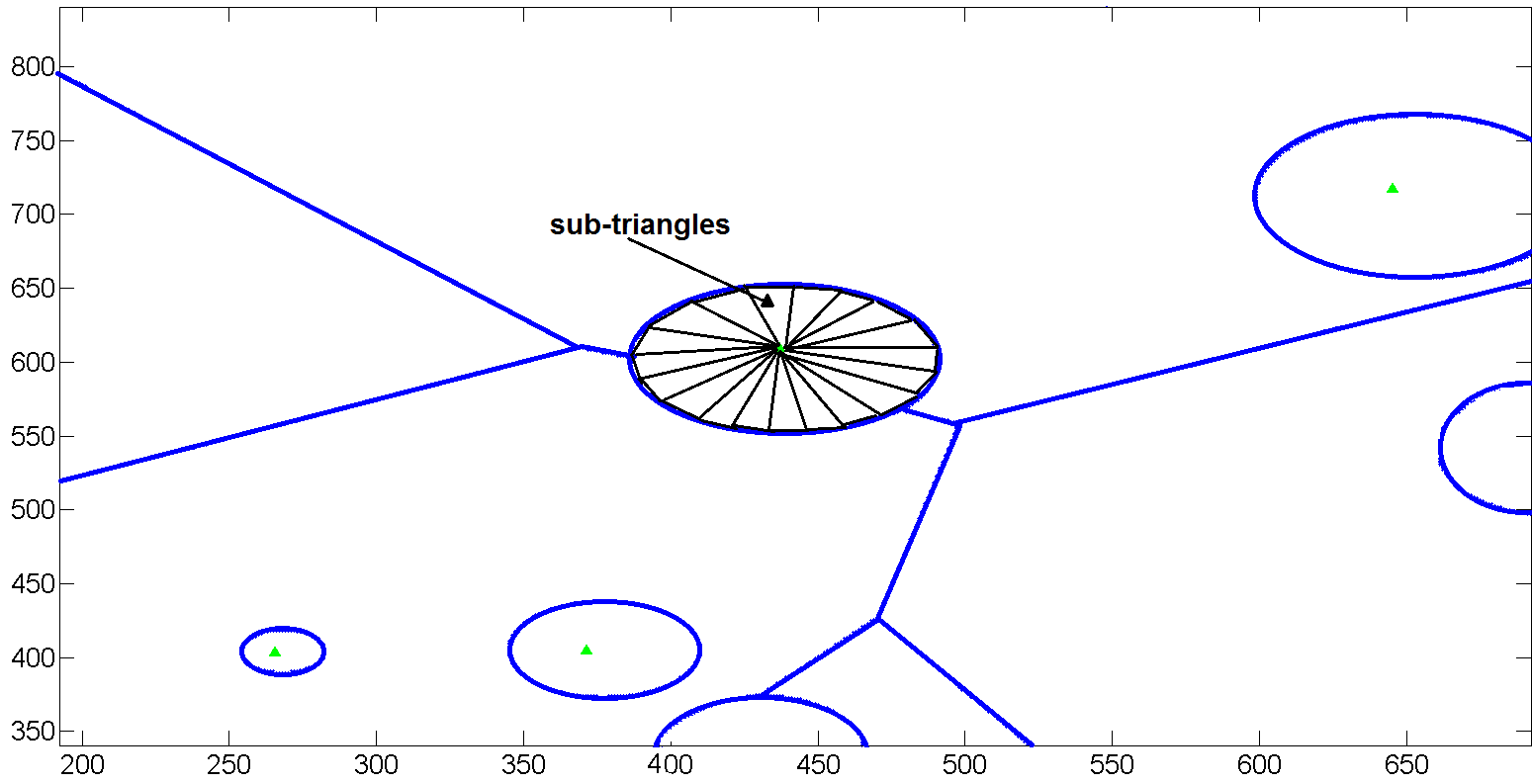}
  \caption{Every BS Voronoi cell can be divided to sub-triangles which are each comprised of the BS as a vertex and the lines connecting the BS to the edge.}
  \label{fig:voronoi_tri}
\end{minipage}
\hspace{5 mm}
\begin{minipage}{.45\textwidth}
  \centering
  \includegraphics[width=\linewidth]{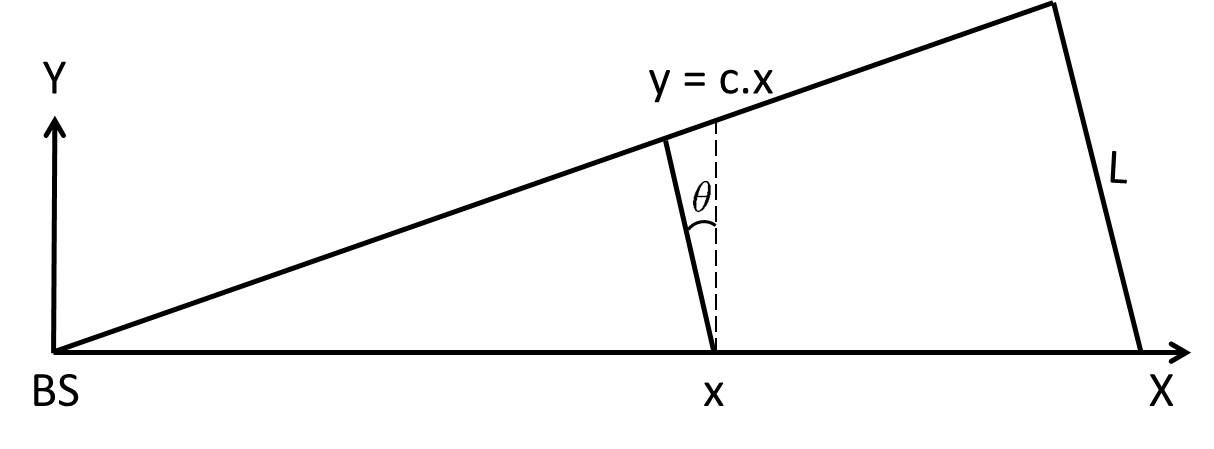}
  \caption{The potential value is assumed to be constant on each line parallel to the cell edges. So, the P value is only a function of $x$.}
  \label{fig:triangle}
\end{minipage}
\end{figure}

First, note that every BS Voronoi cell can be divided to sub-triangles which are each comprised of the BS as a vertex and the lines connecting the BS to the edge. An example is shown in Fig. \ref{fig:voronoi_tri}. Increasing the number of sub-triangles to large numbers assures that the whole area is covered.

Also note that for macro-BSs, some parts of the Voronoi cell might be covered by picocells. However, since the sum of potential value inside each picocell is 0 and the pico-BSs are distributed uniformly, it does not affect the sum potential values in macrocells.

If the rules of potential function hold inside each sub-triangle, we can infer that they hold for entire Voronoi cell because the potential value of each cell is simply the sum of potential values of sub-triangles. Therefore, the required condition for potential function is as follows:
\begin{itemize}
\item $P(x,y) = +1$ for cell center points, 
\item $P(x,y) = -1$ for Voronoi cell edge points,
\item $\iint_{\Delta_{i}}P(x,y)dxdy = 0, \;\; \forall i$,
\end{itemize}
where $\Delta_{i}$ are the sub-triangles.

Since the potential value is assumed to be constant on each line parallel to the cell edges (as shown in Fig. \ref{fig:triangle}), it means that the last condition can be written as follows:
\begin{equation}
\int_{x=0}^{X}\int_{y=0}^{cx}P(x,y)dxdy = \int_{x=0}^{X}cxP(x)dx,
\end{equation}
where constant $c$ is
\begin{equation}
c = \frac{L}{X\text{cos}\theta},
\end{equation}
and $L$ is the length of Voronoi cell edge and $\theta$ is the angle associated with the BS vertex. Increasing the number of sub-triangles, $\theta$ tends to 0 and $\text{cos}\theta$ tends to 1. So,
\begin{equation}
c \simeq \frac{L}{X},
\end{equation}

Now, assume that P is a polynomial function. With substitution of P function in the conditions, we can see that P cannot have a degree of one. So, the minimum degree for P is two. With substitution of a second-degree polynomial function in the potential function conditions, the P function is calculated as
\begin{equation}
\frac{-2x^{2}}{X^2}+1,
\end{equation}
which can be generalized to any arbitrary point $(x,y)$. This proves that the simplest polynomial P function is
\begin{equation}
P(x,y) = \frac{-2(d(x,y))^{2}}{(D(x,y))^{2}}+1.
\end{equation}

\bibliographystyle{IEEEtran}
\bibliography{Meisam_Mirahsan_References}

\begin{IEEEbiography}[{\includegraphics[width=1in,height=1.25in,clip,keepaspectratio]{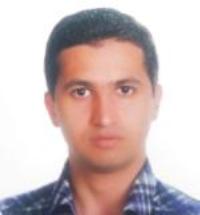}}]{Meisam Mirahsan}
(S'2013) was born in Tehran, Iran in 1983. He received his B.Sc. and M.A.Sc. degrees in Information Technology Engineering and Computer Networks Engineering from Amirkabir University of Technology, Tehran, Iran in 2007 and 2010, respectively.
His research interests include wireless networks, traffic modeling, stochastic geometry, network virtualization, and resource allocation in radio access networks.
He is a Ph.D. candidate in Electrical Engineering at Systems and Computer Engineering department at Carleton University, Ottawa, Ontario, Canada under supervision of Prof. Halim Yanikomeroglu since 2013.
He worked in leading network design company Favapars, Tehran, Iran, 2009-2012 as the network design team manager on Iran's national backhaul IP/MPLS network design.
Since 2015 he is doing an internship as wireless network engineer in Huawei Technologies, Ottawa R\&D, Canada.
\end{IEEEbiography}

\begin{IEEEbiography}[{\includegraphics[width=1in,height=1.25in,clip,keepaspectratio]{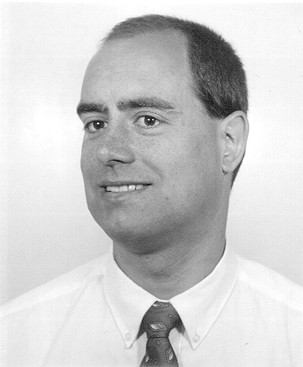}}]{Rainer Schoenen}
(SM'2013) received his German Diplom-Ingenieur and Ph.D.
degrees from RWTH Aachen University,
in Electrical Engineering in 1995 and 2000, respectively.
His research interests include
stochastic Petri nets and queuing systems, ATM,
TCP/IP, switching, flow control, QoS, tariffs,
Userin-the-loop (UIL), wireless resource and packet scheduling
and the MAC layer of 4+5G systems.
His PhD thesis was
“System Components for Broadband Universal Networks with QoS Guarantee.”
with the ISS group of Prof. Heinrich Meyr at RWTH Aachen University, Germany, from 1995 to 2000.
He started working self-employed in 2000.
Dr. Schoenen was a senior researcher at the Communication Networks
(ComNets) Research Group, RWTH Aachen with Professor Walke from 2005 to 2009,
working on computer networks, queuing theory, Petri nets,
LTE-Advanced, FDD relaying, scheduling, OSI layer 2 (MAC) and IMT-Advanced Evaluation within WINNER+.
From 2010 to 2014 he worked at the Department of Systems and Computer Engineering, Carleton University,
Ottawa, Canada, as a project manager supporting Professor Halim Yanikomeroglu.
Since 2015 he is a professor at HAW Hamburg, Germany.
\end{IEEEbiography}

\begin{IEEEbiography}[{\includegraphics[width=1in,height=1.25in,clip,keepaspectratio]{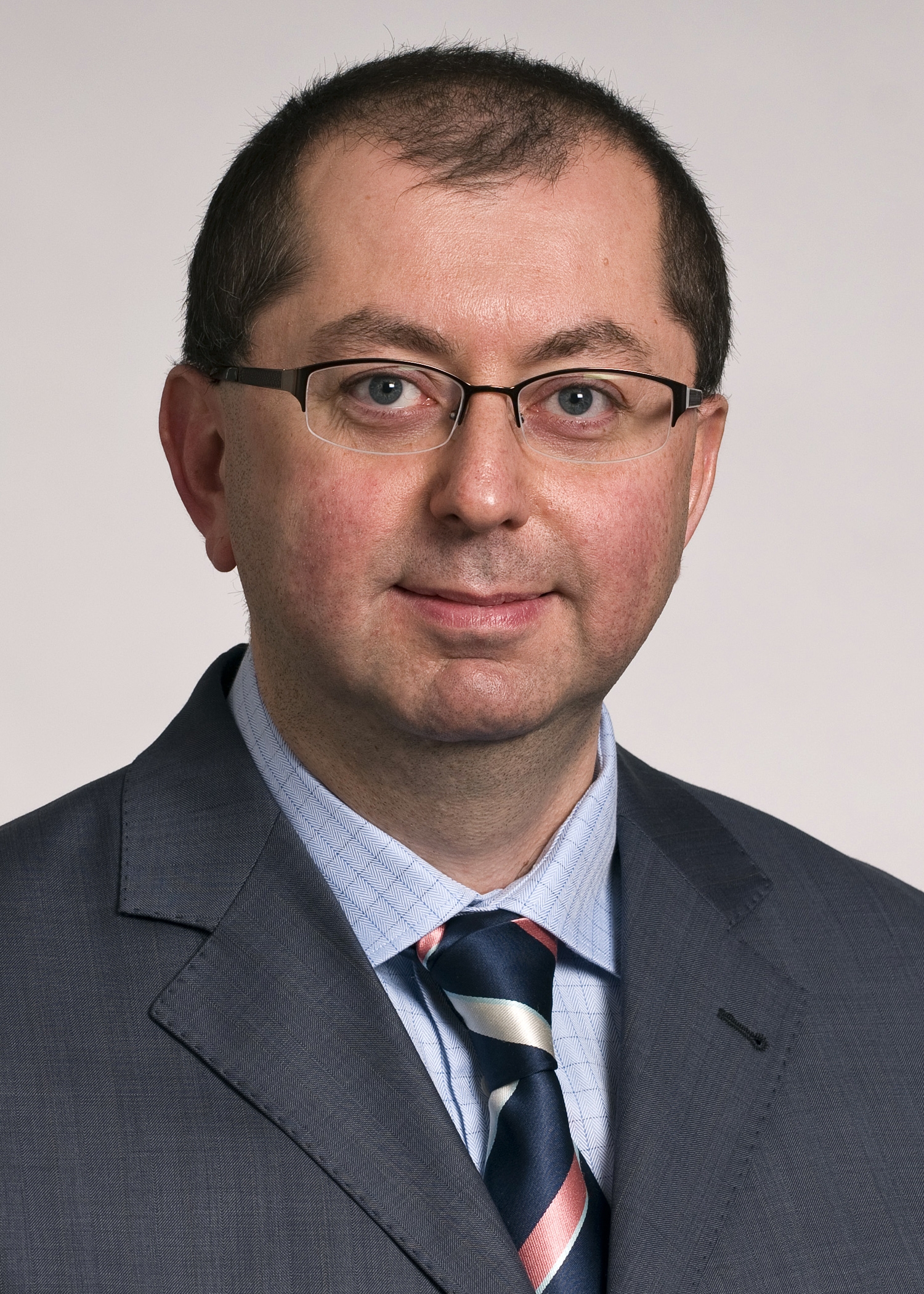}}]{Halim Yanikomeroglu}
(S'96-M'98-SM'12) was born in Giresun, Turkey, in 1968. He received the B.Sc. degree in electrical and electronics engineering from the Middle East Technical University, Ankara, Turkey, in 1990, and the M.A.Sc. degree in electrical engineering (now ECE) and the Ph.D. degree in electrical and computer engineering from the University of Toronto, Canada, in 1992 and 1998, respectively.

During 1993-1994, he was with the R\&D Group of Marconi Kominikasyon A.S., Ankara, Turkey. Since 1998 he has been with the Department of Systems and Computer Engineering at Carleton University, Ottawa, Canada, where he is now a Full Professor. His research interests cover many aspects of wireless technologies with a special emphasis on cellular networks. He coauthored over 60 IEEE journal papers, and has given a high number of tutorials and invited talks on wireless technologies in the leading international conferences. In recent years, his research has been funded by Huawei, Telus, Blackberry, Samsung, Communications Research Centre of Canada (CRC), and Nortel. This collaborative research resulted in about 20 patents (granted and applied). Dr. Yanikomeroglu has been involved in the organization of the IEEE Wireless Communications and Networking Conference (WCNC) from its inception, including serving as Steering Committee Member as well as the Technical Program Chair or Co-Chair of WCNC 2004 (Atlanta), WCNC 2008 (Las Vegas), and WCNC 2014 (Istanbul). He was the General Co-Chair of the IEEE Vehicular Technology Conference Fall 2010 held in Ottawa. He has served in the editorial boards of the IEEE TRANSACTIONS ON COMMUNICATIONS, IEEE TRANSACTIONS ON WIRELESS COMMUNICATIONS, and IEEE COMMUNICATIONS SURVEYS \& TUTORIALS. He was the Chair of the IEEE’s Technical Committee on Personal Communications (now called Wireless Technical Committee). He is a Distinguished Lecturer for the IEEE Communications Society (2015-2016) and the IEEE Vehicular Technology Society (2011-2015).

Dr. Yanikomeroglu is a recipient of the IEEE Ottawa Section Outstanding Educator Award in 2014, Carleton University Faculty Graduate Mentoring Award in 2010, the Carleton University Graduate Students Association Excellence Award in Graduate Teaching in 2010, and the Carleton University Research Achievement Award in 2009. Dr. Yanikomeroglu spent the 2011–2012 academic year at TOBB University of Economics and Technology, Ankara, Turkey, as a Visiting Professor. He is a registered Professional Engineer in the province of Ontario, Canada.
\end{IEEEbiography}

\end{document}